\documentclass[aps,prb,twocolumn,showpacs,showkeys,groupedaddress]{revtex4-2}

\usepackage{amsmath}
\usepackage{graphicx}
\usepackage{bm}
\usepackage{xcolor}
\usepackage{epstopdf}

\usepackage{hyperref}
\usepackage[all]{hypcap}

\renewcommand{\d}{\ensuremath{\mathrm{d}}}

\begin{document}
\title{Superconducting Diode Effect in Quantum Spin Hall Insulator-based Josephson Junctions}

\author{Benedikt Scharf}

\affiliation{Institute for Theoretical Physics and Astrophysics and W\"{u}rzburg-Dresden Cluster of Excellence ct.qmats, University of W\"{u}rzburg, Am Hubland, 97074 W\"{u}rzburg, Germany}

\author{Denis Kochan}
\email{corresponding author: denis.kochan@savba.sk}

\affiliation{Institute of Physics, Slovak Academy of Sciences, 84511 Bratislava, Slovakia}
\affiliation{Center for Quantum Frontiers of Research and Technology (QFort), National Cheng Kung University, Tainan 70101, Taiwan}

\author{Alex Matos-Abiague}

\affiliation{Department of Physics $\&$ Astronomy, Wayne State University, Detroit, MI 48201, USA}

\date{\today}
 
\begin{abstract}
The superconducting diode effect (SDE) is a magneto-electric phenomenon where an external magnetic field imparts a non-zero center-of-mass momentum to Cooper pairs, either facilitating or hindering the flow of supercurrent depending on its direction. We propose that quantum spin Hall insulator (QSHI)-based Josephson junctions can serve as versatile platforms for non-dissipative electronics exhibiting the SDE when triggered by a phase bias and an out-of-plane magnetic field.
By computing the contributions from Andreev bound states and the continuum of quasi-particle states, we provide both numerical and analytical results scrutinizing various aspects of the SDE, including its quality Q-factor. The maximum value of the $Q$-factor is found to be universal at low (zero) temperatures, which ties its origin to underlying topological properties that are independent of the junction's specific details. As the magnetic field increases, the SDE diminishes due to the closing of the induced superconducting gap caused by orbital effects.
To observe the SDE, the QSHI-based Josephson junction must be designed so that its edges are transport-wise non-equivalent. Additionally, we explore the SDE in a more exotic yet realistic scenario, where the fermionic ground-state parity of the Josephson junction remains conserved while driving a current. In this 4$\pi$-periodic situation, we predict an enhancement of the SDE compared to its 2$\pi$-periodic, parity-unconstrained counterpart.
\end{abstract}

\maketitle

\section{Introduction}

Magneto-electric phenomena accompanying superconductors with broken 
time-reversal and space-inversion symmetries are attracting considerable attention.\cite{Nadeem_NatRevPhys2023} There are numerous superconducting systems displaying 
the supercurrent diode effect (SDE) including 
(1) thin superconducting films,\cite{Sivakov2018:LowTempPhys-Exp,Vodolazov2018-SuperScinTech,Ando2020, Hou2023:PRL} 
(2) Josephson junctions (JJs) based on 
(i) semiconductors,\cite{Baumgartner2022,BaumgartnerSI2022,Turini2022,Costa2023:NatureNanotechnology,Lotfizadeh:CP2024,Reinhardt2024:NatureCommm}
(ii) topological semimetals,\cite{Pal2022}
(iii) proximity-magnetized metals with strong spin-orbit coupling (SOC),\cite{Jeon2022} 
(iv) van der Waals heterostructures,\cite{Wakatsuki2017,Wu2022,Bauriedl2022}
(v) twisted bilayer\cite{DiezMerida2023:NatComm} 
and trilayer\cite{Lin2022:NatPhys-experiment} graphenes, 
(vi) high-$T_c$ superconductors,\cite{Zhao2023:S,Ghosh2024:NM}
(vii) ferromagnets,\cite{Pal2019:EPL}
(viii) topological insulators\cite{Chen2018:PRB,Tanaka2022:PRB,Lu2023:PRL,Fu2024:PRApp} or semiconductor-based Majorana wires,\cite{Legg2023:PRB,Cayao2024:PRB,Meyer2024:APL}
(ix) spin-orbit coupled quantum dot junctions,\cite{Debnath2024:PRB}
(3) Josephson weak links through a single magnetic atom\cite{Trahms2023:Nature}
or even
(4) altermagnets\cite{banerjee2024altermagnetic,Zhang-Neupert2024:NatComm}.
As many of them demonstrate the potential for supercurrent rectification, that is, maintaining a system superconducting for one current direction, while transiting it to the resistive state for the opposite one, they digress in the roles played by magnetic fields, Meissner screening, magnetization, the origin of spin-momentum locking, and generically in the spin-resolved spectral properties of the associated subgap states. 
All these nuances diversify the SDE and pin its origin with various proliferating adjectives like 
intrinsic, 
trivial,
universal,\cite{Davydova2022} 
anomalous,\cite{Fracassi2024:APL}
ubiquitous,\cite{Hou2023:PRL},
field-free,\cite{Wu2022} 
single-atomic,\cite{Trahms2023:Nature}   
flux-tunable,\cite{Coraiola:ASCNano2024}
high-temperature,\cite{Ghosh2024:NM}
altermagnetic\cite{banerjee2024altermagnetic}
and transverse,\cite{Fu2024:arxiv} among others.

There is a common agreement linking the SDE to the appearance of a finite center-of-mass momentum of Cooper pairs, although, there is less consensus on what causes its non-zero value. The breakdown of time-reversal 
symmetry, triggered by a magnetic field (via Zeeman coupling) or magnetization (via exchange splitting), and a 
moving condensate while probing the supercurrent seem to be necessary ingredients in all scenarios.\cite{Edelstein1989,Edelstein1996,Daido2021,Yuan2021,Smith2021,He2022,Scammell2022,Ilic2022,Davydova2022,PicoliPRB:2023-Theory,Baumgartner2022,BaumgartnerSI2022,Fuchs2022,Costa2023:NatureNanotechnology,Hou2023:PRL,Banerjee2023phase,Sundaresh2023:NatCom,Lotfizadeh:CP2024,Reinhardt2024:NatureCommm}
However, there is still an ongoing discussion to which extent the SOC plays a role. Since experiments vary in terms of materials, geometry, measurements, and even the nature of superconductivity (whether intrinsic or proximity-induced), no single theory can comprehensively explain the SDE in all its various forms.
The primary distinction between models of the SDE lies in their stance on SOC. Pro-SOC models\cite{Edelstein1989,Edelstein1996,Daido2021,Yuan2021,Smith2021,He2022,Scammell2022,Ilic2022,Baumgartner2022,BaumgartnerSI2022,Costa2023:NatureNanotechnology,Kochan2023diode,Costa2023:PRB-Theory,Lotfizadeh:CP2024} link supercurrent rectification to the emergence of a helical superconducting phase.\cite{Edelstein1989,Mineev1994,Edelstein1996,Dimitrova2007,Buzdin2008,Mineev2012} In contrast, con-SOC models\cite{Davydova2022,PicoliPRB:2023-Theory,Hou2023:PRL,Sundaresh2023:NatCom,Banerjee2023phase} attribute the SDE either to a Doppler shift in the quasi-particle spectra for left and right movers or to diamagnetic effects resulting from stray fields and inhomogeneous screening, or alternatively to 
Yu-Shiba-Rusinov states\cite{Trahms2023:Nature,Costa2018}.

In this paper we scrutinize magneto-chiral properties of two-dimensional (2D) JJs based on Quantum Spin Hall Insulators (QSHI) and explore their abilities to foster the SDE when tuning an out-of-plane magnetic field, the phase biasing, and/or the edge-channels asymmetry. 
We compute individual contributions to the supercurrent carried by the spin-resolved Andreev bound states (ABS) and the continuum quasi-particle states, and analyze their roles in the emergence of the SDE.
Using a minimal yet realistic model capturing the main junction characteristics and considering the relevant system parameters, we investigate the SDE and provide practical analytical formulas for the underlying quality Q-factor, focusing primarily on its dependencies on magnetic field and temperature. We show that in the presence of an out-of-plane magnetic field and at very low temperatures the maximum of the $Q$-factor acquires a universal value, which is independent of the junction parameters. Finally, we show that the SDE and its corresponding $Q$-factor can be further enhanced by considering $4\pi$-periodic junctions where the fermionic ground-state parity remains conserved while probing the supercurrent, which requires a measurement of the SDE on timescales shorter than the quasi-particle poisoning time.

The paper is organized as follows: In 
Sec.~\ref{Sec:Model} we present the model Hamiltonian and discuss the associated Andreev and continuous spectra and their corresponding density of states. 
In Sec.~\ref{Sec:FreeEnergyCurrent} we study the underlying free energy and Josephson current and their dependencies on temperature and magnetic field, among other junction parameters.
The effects of the magnetic field, edge-state Fermi velocities, and temperature on the properties of the SDE and $Q$-factor in QSHI-based 2D junctions are analyzed in Sec.~\ref{Sec:Qfactor}. Section~\ref{Sec:Parity} is devoted to the discussion of the SDE when the fermionic ground-state parity is conserved.
We examine possible ramifications toward experimental realizations of the SDE in QSHI-based JJs in Sec.~\ref{Sec:ExpReal}. 
After the concluding remarks, the paper is supplemented with two brief technical Appendices. 

We use standard notations throughout this paper, where $e>0$ stands for the elementary charge, $\Phi_0=h/e$ is the quantum of the magnetic flux, and $k_{\text{B}}$ is the Boltzmann constant. Additionally, $\Theta(x)$ and 
$\text{sgn}(x)$ denote the Heaviside step function and sign function, respectively. 

\begin{figure}[t]
\centering
\includegraphics*[width=8.5cm, trim={0 7cm 0 0},clip]{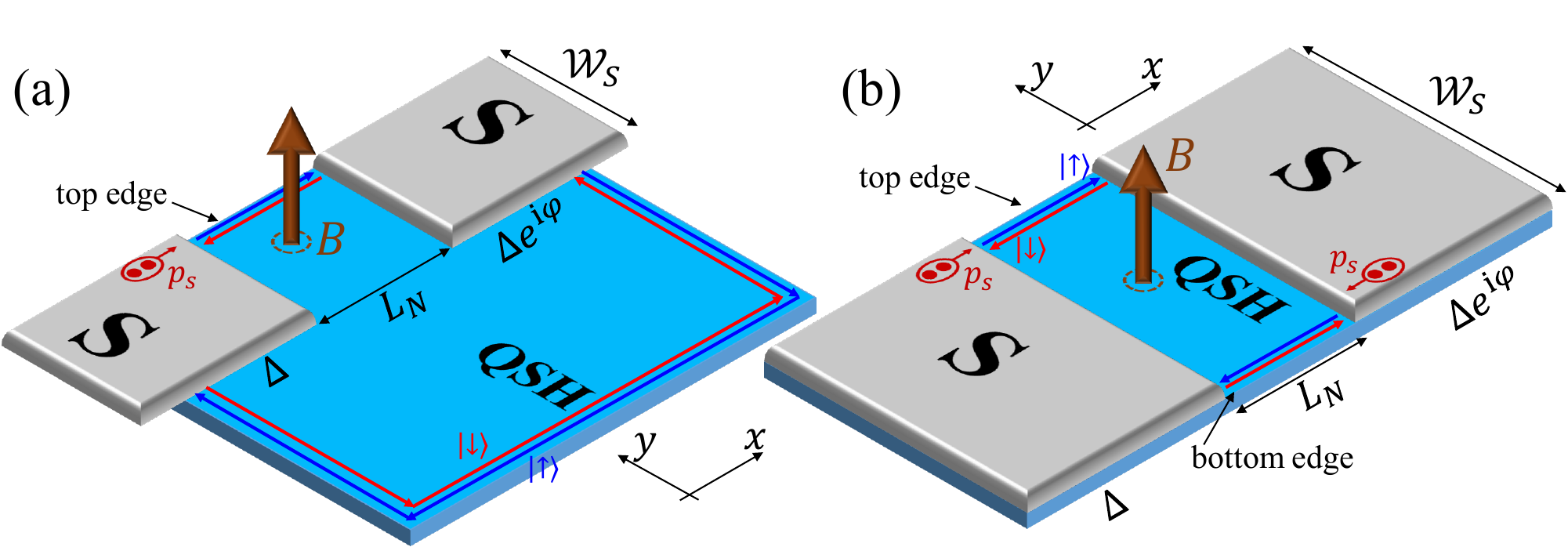}
\caption{Topological Josephson junctions formed by (a) one and (b) two QSHI edges. Here the QSHI is partially covered by two $s$-wave superconductors with different superconducting phases, the Cooper-pair tunneling induces superconducting proximity pairings in the corresponding regions of the QSHI, defining the superconducting (S) and normal (N) spacers. 
The N region is subtended by a perpendicular magnetic field $\bm{B}$. The coordinate system is chosen in such a way that the QSHI lies in the $xy$ plane, with $x$ being the transport direction defined by the gradient of the superconducting phase, correspondingly, $\bm{B}=B\mathbf{e}_z$. 
The edge with a positive (negative) ordinate $y$ is called the top (bottom).}\label{fig:Setup}
\end{figure}

\section{Model, Andreev bound states and continuum density of states}\label{Sec:Model}

We consider a short QSHI-based topological JJ\cite{Dolcini2015:PRB,Scharf2021:PRR,Tkachov2015:PRB} employing the established $\delta$-function model developed in Ref.~\onlinecite{Scharf2021:PRR} [for the geometry and configuration see Fig.~\ref{fig:Setup}, where for the most part of the paper we focus on the setup with a single proximitzed edge as shown in Fig.~\ref{fig:Setup}(a)]. 
We assume the Fermi level is tuned to the bulk gap of the QSHI. Hence the only propagating modes are the spin-polarized edge states. When the width $\mathcal{W}_S$ of the junction is wide enough ---as assumed in this paper--- the QSHI edge states (referred to as the top and bottom edges/channels) do not overlap and their low-energy dynamics decouple. Correspondingly, the physics at each edge can be accurately described by the Bogoliubov-de Gennes (BdG) Hamiltonian,
\begin{multline}\label{eq:BDGHamSimple}
\hat{H}^\sigma_s=\left(s\sigma v_F\hat{p}_x-\mu_S\right)\tau_z+s\frac{v_Fp_S}{2}+V_0h(x)\tau_z\\
+\Delta\left[\tau_x\cos\Phi^\sigma(x)-\tau_y\sin\Phi^\sigma(x)\right].
\end{multline}
In the above equation, $\hat{p}_x=-i\hbar\partial_x$ denotes the momentum operator, $\mu_S$ represents the chemical potential in the superconducting~(S) regions,$V_0$ is the potential difference between the normal~(N) and S regions, and the (super/sub)scripts $s=\uparrow\hspace{-1mm}/\hspace{-1mm}\downarrow\equiv\pm1$ and $\sigma=t/b\equiv\pm1$ represent the spin projections along the $z$-axis and the top and bottom channels, respectively. Finally $\tau_{x,y,z}$ are Pauli matrices operating on the particle-hole degrees of freedom. We consider junctions where the top and bottom edge states have the same Fermi velocity ($v_F=v_F^{t/b}$), although junctions with $v_F^t\neq v_F^b$ are also possible, as discussed in Sec.~\ref{Sec:ExpReal}. We limit our study to short junctions, where the length of the N region $L_N$ is shorter than the superconductor coherence length. Furthermore, we also assume $L_N$ to be shorter than the Josephson penetration depth, such that Josephson vortices and Fraunhofer features do not affect the system under consideration. Therefore, the N region can be effectively modelled as a $\delta$-like spacer with a potential profile $h(x)=L_N\delta(x)$.
The QSHI is proximitized by two $s$-wave superconductors which are assumed to have equal gap magnitudes, 
but different superconducting phases. Consequently, the induced superconducting pairing in the QSHI, 
$\Delta e^{i\phi(x)}$, retains a constant gap amplitude $\Delta$, but an $x$-dependent phase $\phi(x)$, $\phi(x<0)=0$ and $\phi(x>0)=\phi$, resulting in a global phase difference $\phi$. 
An out-of-plane magnetic field $\bm{B}=B\bm{e}_z$ with $B\geq0$ induces:\cite{Tkachov2015:PRB} (i)~an orbital Doppler shift described by the Cooper pair momentum 
$p_S$ and (ii)~the edge-selective superconducting-phase profiles $\Phi^\sigma(x)$ with~$\Phi^\sigma(x<0)=0$ and $\Phi^\sigma(x>0)=\phi^\sigma$, where
\begin{equation}\label{eq:DopplerOnes}
    p_S=\pi\hbar\,\frac{B\,\mathcal{W}_S}{\Phi_0}\,,\ \ \ \ \ 
    \phi^\sigma=\phi+\sigma\,\frac{p_S\,L_N}{\hbar}\,.
\end{equation}
The second term in $\phi^\sigma$ can be interpreted as a kinematic phase acquired by a Cooper pair with a momentum $\sigma p_S$ when traversing a distance of $L_N$.

\begin{figure}[t]
\centering
\includegraphics*[width=8.5cm]{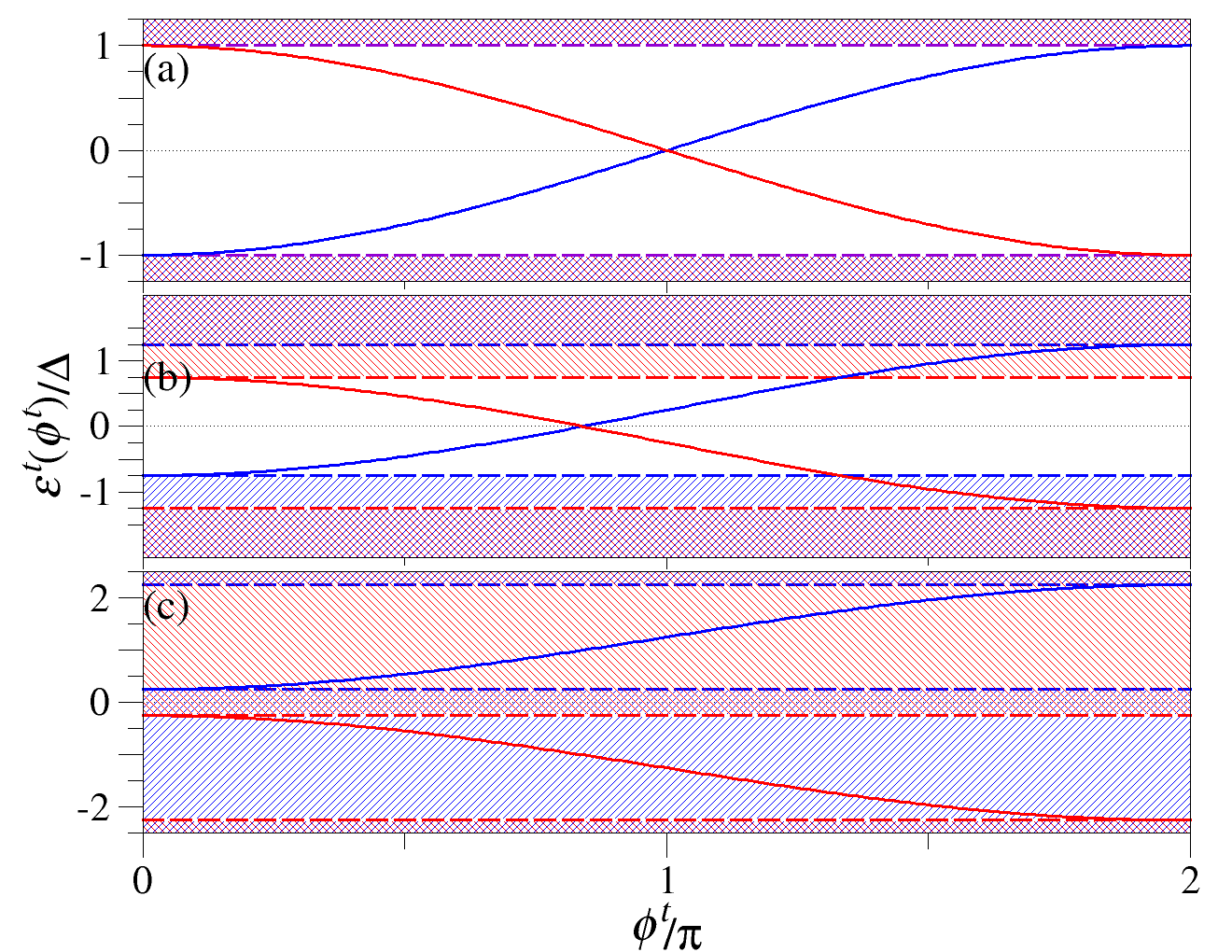}
\caption{ABS given by Eq.~(\ref{eq:ABS}) for the top edge for different values of the dimensionless parameter $\gamma=v_Fp_S/(2\Delta)$ [defined below in Eq.~(\ref{eq:I0gamma})]: (a) $\gamma=0$, (b) $\gamma=0.25$, (c) $\gamma=1.25$. Here the blue and red colors denote different spin 
quantum numbers $s=\uparrow\hspace{-1mm}/\hspace{-1mm}\downarrow\equiv\pm1$ associated with the sub-gap 
ABS (solid lines) and the supra-gap continuum states (shadings).}\label{fig:ABS}
\end{figure}

Following Refs.~\onlinecite{Scharf2021:PRR,Tkachov2019:PRB}, the ABS spectra associated with 
the model Hamiltonian, Eq.~(\ref{eq:BDGHamSimple}), are determined by means of the spectral scattering approach, which yields per each edge (indexed by $\sigma$) two ABS branches (indexed by spin $s$) 
with an obvious $2\pi$ periodicity,
\begin{equation}\label{eq:ABS}
\epsilon^\sigma_s(\phi^\sigma)=s
\Bigl[-\sigma\Delta\cos\frac{\phi^\sigma}{2}\mathrm{sgn}
\Bigl(
\sin\frac{\phi^\sigma}{2}
\Bigr)+\frac{v_Fp_S}{2}
\Bigr].
\end{equation}
When $\phi^\sigma=\phi^{t/b}$ approaches an integer multiple of $2\pi$, a pair of the sub-gap ABS with opposite spin projections merges into the continuum of quasi-particle states, see Fig.~\ref{fig:ABS}, while 
a pair of ABS with the reversed spins splits-off from the supra-gap states. 

Complementary, the continuum of quasi-particles contributes to the supra-gap density of states (DOS) that can be separated into a phase-independent (shown explicitly later) and phase-dependent part. 
It is the latter, which is important for the computation of the Josephson current via differentiation of the free energy with respect to the superconducting phase difference. The phase-dependent part of the DOS, per 
spin and edge, can be compactly written as,
\begin{equation}\label{eq:ContinuumDOS}
{\rho}^\sigma_s(\epsilon,\phi^\sigma)=\frac{s\sigma\Delta^2}{2\pi}\frac{\Theta\left(\epsilon_s^2-\Delta^2\right)\mathrm{sgn}\left(\epsilon_s\right)\,\sin\phi^\sigma}{\sqrt{\epsilon_s^2-\Delta^2}\bigl(\epsilon_s^2-\Delta^2\cos^2\frac{\phi^\sigma}{2}\bigr)}
\end{equation}
with 
\begin{equation}
    \epsilon_{s}=\epsilon -s v_Fp_S/2\,.
\end{equation}
Although ${\rho}^\sigma_s(\epsilon,\phi^\sigma)$ can be negative, the total DOS, which also includes the 
phase-independent part, is always positive. 
As discussed in Ref.~\onlinecite{Scharf2021:PRR}, a finite $v_Fp_S$ decreases the effective superconducting gap, which closes for $|v_Fp_S|\geq2\Delta$. However, for each spin species saddled 
by the opposite edges of the QSHI the superconducting gaps still remain open individually, and although energetically not overlapping, see Fig.~\ref{fig:ABS}(c), the bound and continuum states exist simultaneously on a spatial scale defined by the junction width $\mathcal{W}_S$. In the following, we will focus mainly on the setup with only one proximitized edge, as shown in Fig.~\ref{fig:Setup}(a).

\section{Free energy and current}\label{Sec:FreeEnergyCurrent}

The ABS spectra, Eq.~(\ref{eq:ABS}), and the supra-gap DOS due to continuum states, Eq.~(\ref{eq:ContinuumDOS}), allow us to determine the free energy from which the Josephson current and other thermodynamic quantities follow straightforwardly. As the edges of the QSHI-based JJ are dynamically decoupled, the total free energy turns into a sum of its top and bottom parts. Hence, in an equilibrium held 
at temperature $T$ and magnetic field $B\geq 0$ the free energy $F^\sigma(\phi^\sigma,T)$ of the edge $\sigma$ (more precisely its phase-dependent part) equals,
\begin{multline}\label{eq:FreeEnergy}
F^\sigma(\phi^\sigma,T)=-k_{\text{B}}T
\biggl\{
\ln
\biggl[
2\cosh
\biggl(
\frac{\epsilon^\sigma_\uparrow(\phi^\sigma)}{2k_{\text{B}}T}
\biggr)
\biggr]
\\
+\int\limits_0^\infty\d\epsilon\,\rho^\sigma(\epsilon,\phi^\sigma)\,\ln\left[2\cosh\left(\frac{\epsilon}{2k_{\text{B}}T}\right)\right]
\biggr\}.
\end{multline}
Equation~(\ref{eq:FreeEnergy}) consists of a discrete part due to ABS and of a continuum part, which integrates over the supra-gap DOS summed over both spin projections,
\begin{equation}\label{eq:DOS}
\rho^\sigma(\epsilon,\phi^\sigma)=\sum_{s=\uparrow/\downarrow}\rho^\sigma_s(\epsilon,\phi^\sigma).
\end{equation}

\begin{figure}[t]
\centering
\includegraphics*[width=8.5cm]{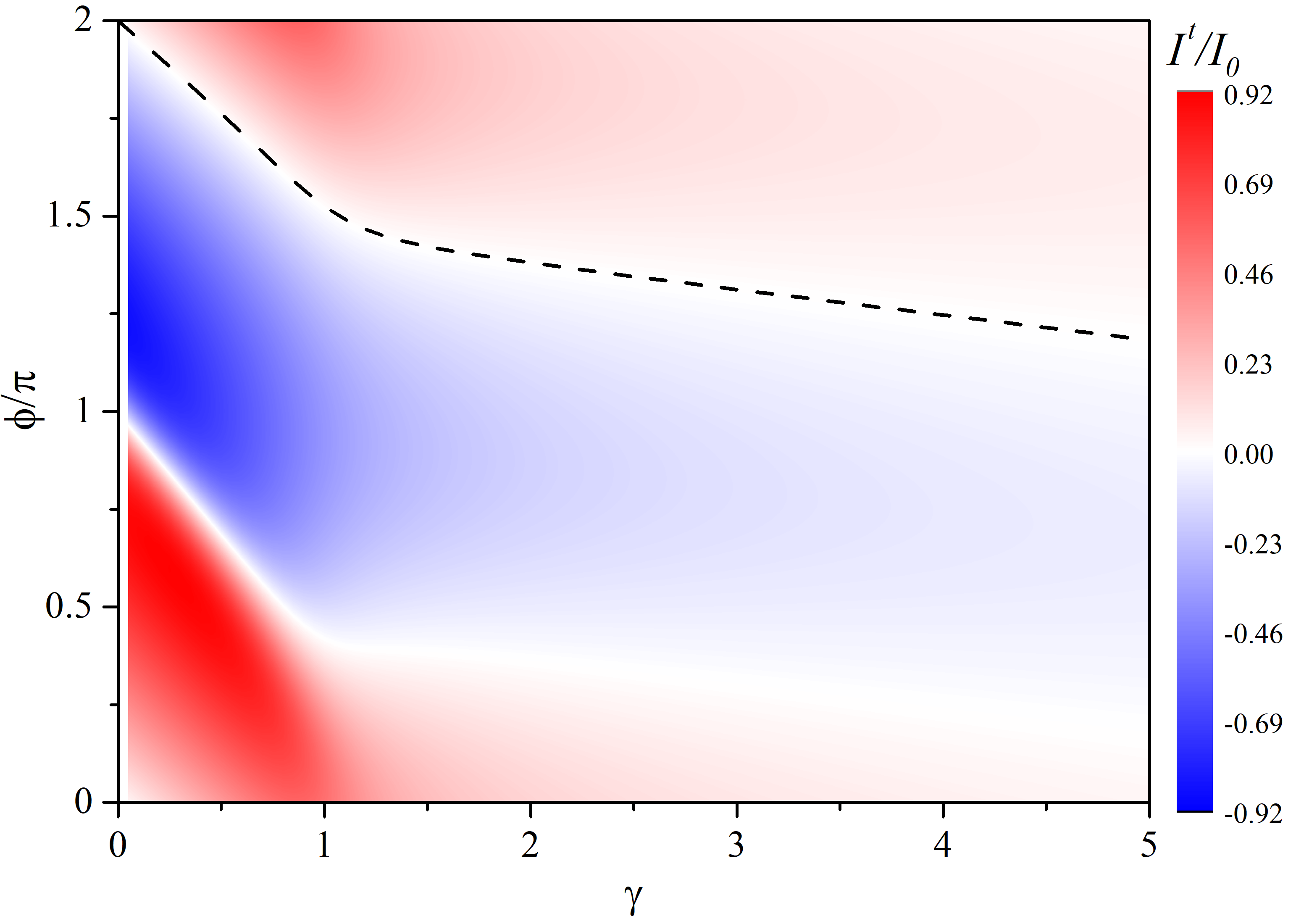}
\caption{Josephson current $I^t$, composed of the ABS and continuum states, carried by the top 
edge of the QSHI-based junction plotted as a function of the phase difference $\phi$ and parameter $\gamma=v_Fp_S/(2\Delta)\propto B$. The dashed line traces the evolution of the superconducting phase difference $\phi$ that minimizes the top-edge free energy at a given value of $\gamma$.
Here $k_{\text{B}}T=0.1\Delta$, $L_N\Delta/\hbar v_F=0.1$ and the current $I^t$ is measured in units of $I_0=e\Delta/(2\hbar)$.}\label{fig:JosephsonCurrentTop3D}
\end{figure}

As the superconducting phase difference $\phi$ and the edge-saddled phase difference $\phi^\sigma$ differ by just a shift, Eq.~(\ref{eq:DopplerOnes}), the Josephson current carried by an edge $\sigma$ is then calculated as,
\begin{equation}\label{eq:JosephsonCurrentDef}
I^\sigma(\phi^\sigma,T)=\frac{2e}{\hbar}\frac{\partial F^\sigma(\phi^\sigma,T)}{\partial\phi^\sigma}
=\frac{2e}{\hbar}\frac{\partial F^\sigma(\phi^\sigma,T)}{\partial\phi}.
\end{equation}
Inserting Eqs.~(\ref{eq:ABS})~and~(\ref{eq:ContinuumDOS}) into the
free energy $F^\sigma(\phi^\sigma,T)$, the Josephson current splits
into ABS and continuum contributions,
\begin{equation}\label{eq:JosephsonCurrent}
I^\sigma(\phi^\sigma,T)=I_{\text{ABS}}^{\sigma}(\phi^\sigma,T)+I_{\text{cont}}^{\sigma}(\phi^\sigma,T)\,.
\end{equation}
Introducing the short-hand notations,
\begin{equation}\label{eq:I0gamma}
I_0=\frac{e\Delta}{2\hbar},\ \ \ \ \ \gamma=\frac{v_F p_S}{2\Delta},\ \ \ \ \ \tilde{\Delta}=\frac{\Delta}{2k_{\text{B}} T},
\end{equation}
the current contributions read,
\begin{multline}\label{eq:JosephsonCurrentABS}
I_{\text{ABS}}^\sigma(\phi^\sigma,T)=
I_0\sin\frac{\phi^\sigma}{2}\quad\quad\\
\times
\tanh\Bigl[\tilde{\Delta}\Bigl(\cos\frac{\phi^\sigma}{2}-\sigma\gamma\,\mathrm{sgn}\Bigl(\sin\frac{\phi^\sigma}{2}\Bigr)\Bigl)\Bigr]\,
\end{multline}
and
\begin{multline}\label{eq:JosephsonCurrentContinuum}
I_{\text{cont}}^\sigma(\phi^\sigma,T)=-I_0\,\frac{\sigma}{\tilde{\Delta}}\,\frac{1}{\pi}\int\limits_1^{\infty}\d x\;
\mathrm{ln}
\biggl[
\frac{\cosh\bigl(\tilde{\Delta}(x+\gamma)\bigr)}{\cosh\bigl(\tilde{\Delta}(x-\gamma)\bigr)}
\biggr]\\
\times\frac{x^2\cos\phi^\sigma-\cos^2\frac{\phi^\sigma}{2}}{\sqrt{x^2-1}
\bigl(x^2-\cos^2\frac{\phi^\sigma}{2}\bigr)^2}\,.
\end{multline}
At zero magnetic field, $\gamma$ reduces to zero and consequently Eq.~(\ref{eq:JosephsonCurrentContinuum}) gives no contribution, which is consistent with the expectation that in field-free short junctions the supercurrent is driven by the sub-gap states.

\begin{figure}[t]
\centering
\includegraphics*[width=8.5cm]{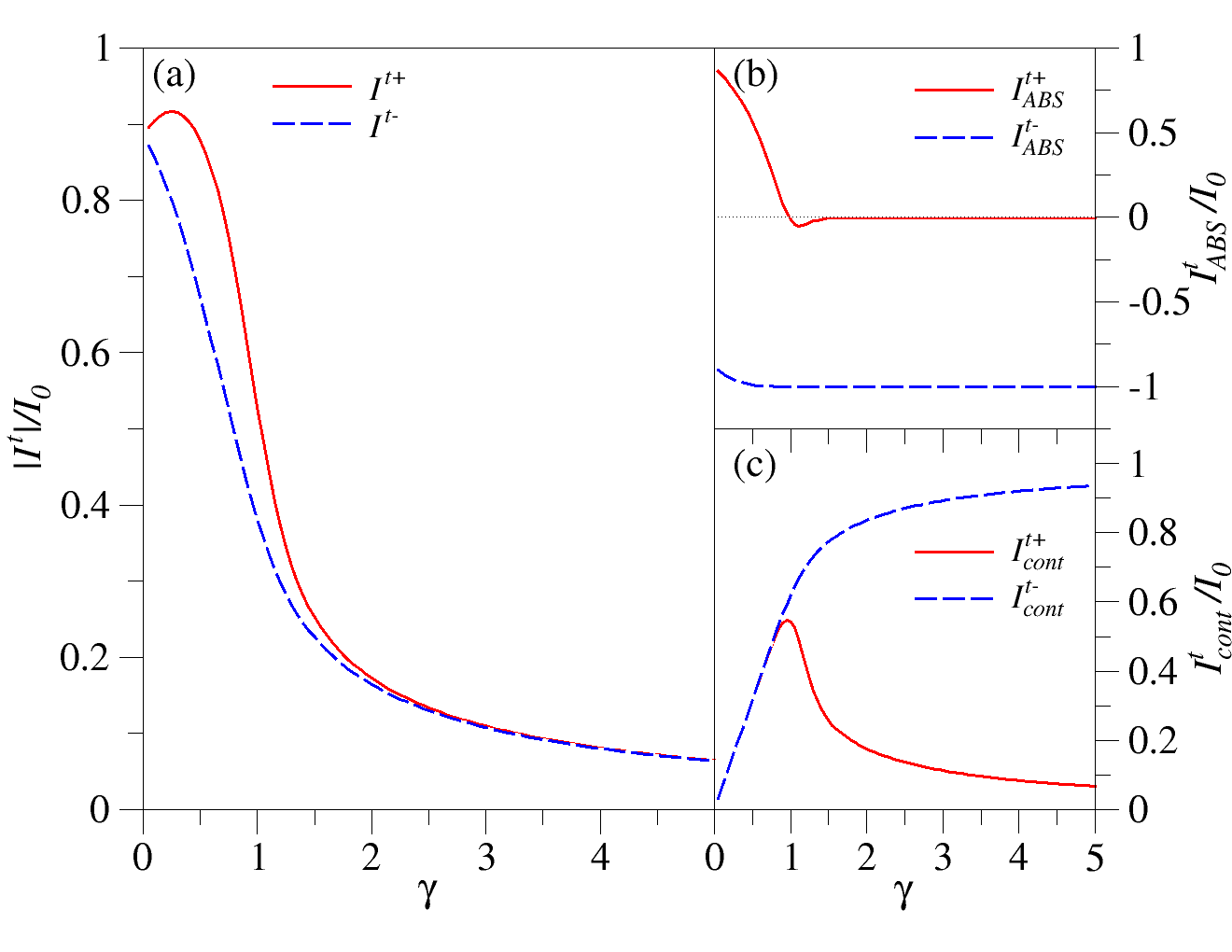}
\caption{Top edge Josephson current characteristics for a QSHI-based JJ. Panel~(a) displays maximal (red), $I^{t +}$, and minimal (blue), $I^{t -}$, critical currents (normalized to $I_0$) as functions of $\gamma=v_Fp_S/(2\Delta)\propto B$. 
Panels~(b)~and~(c) show, correspondingly, the ABS contribution, $I_{\text{ABS}}^{t\pm}$, and continuum-state contribution, $I_{\text{cont}}^{t\pm}$, of $I^{t \pm}$. Again the maxima and minima are plotted in red and blue.}\label{fig:JosephsonCurrentTop}
\end{figure}

Figure~\ref{fig:JosephsonCurrentTop3D} shows the top-edge current-phase relation, $I^t(\phi)$,
normalized to $I_0$ at $k_{\text{B}}T=0.1\Delta$ as a function of the phase difference $\phi$ and the dimensionless parameter $\gamma\propto B$ [see Eq.~(\ref{eq:I0gamma})]. 
As $\gamma$ increases, the maxima (red loci), minima (blue loci) and zeros (white trench) of the current-phase relation move along the $\phi$ axis, while, simultaneously, the amplitude of the Josephson current reduces.

It follows from Eq.~(\ref{eq:JosephsonCurrentDef}), that the points $(\gamma,\phi^t)$ where $I^t$ vanishes and the free energy $F^t$ minimizes---shown in Fig.~\ref{fig:JosephsonCurrentTop3D} by the black-dashed line---determine
the ground-state phase, i.e., the phase at which the top edge is in its ground state for a given $\gamma$ (magnetic field). As shown in Fig.~\ref{fig:JosephsonCurrentTop3D}, increasing $\gamma$ causes the ground-state phase to shift from $2\pi (\sim 0$) to $\pi$, resulting in the top junction edge undergoing a $0-\pi$-like transition. The second white trench seen in Fig.~\ref{fig:JosephsonCurrentTop3D}, emanating from $\phi=\pi$ at $\gamma=0$, corresponds to maxima of the free energy $F^t$. The supercurrent at the bottom edge can readily be obtained from Fig.~\ref{fig:JosephsonCurrentTop3D} by using the symmetry relation $I^b(\phi,\gamma)=-I^t(2\pi-\phi,\gamma)$.

In the low-temperature limit, the integration in Eq.~(\ref{eq:JosephsonCurrentContinuum}) can be performed analytically and the contribution from the continuum of states at the top/bottom edge reduces to
\begin{widetext}
\begin{equation}\label{eq:JosephsonCurrentContinuumApprox}
I_{\text{cont}}^\sigma(\phi^\sigma,T\to0)=\sigma\frac{2 I_0}{\pi}
\Bigl\{
\gamma+\Theta(\gamma-1)
\Bigl[
\arctan
\Bigl(
\frac{\sqrt{\gamma^2-1}}{\sin\frac{\phi^\sigma}{2}}
\Bigr)\sin\frac{\phi^\sigma}{2}-\sqrt{\gamma^2-1}
\Bigr]
\Bigr\}.
\end{equation}
\end{widetext}

Note that the above equation has been derived assuming $B\geq0$ or equivalently $\gamma\geq0$. To obtain the corresponding expression for $B<0$, the relation $I^{\sigma}(B,\phi)=I^{-\sigma}(-B,\phi)$ 
can be used. It is instructive to connect the dependence of $I_{\text{cont}}^\sigma(\phi^\sigma,T\to0)$ on $\gamma$ with the closing of the effective superconducting gap, see Fig.~\ref{fig:ABS}(b), and of the lifting of the protected ABS crossing, see Fig.~\ref{fig:ABS}(c), when $\gamma=v_F p_S/(2\Delta)$ exceeds unity.

\section{Superconducting diode effect and $Q$-factor}\label{Sec:Qfactor}

The maximum $I^{\sigma +}=\mathrm{max}_{\phi^\sigma}I^\sigma(\phi^\sigma)$ and minimum $I^{\sigma -}=\mathrm{min}_{\phi^\sigma}I^\sigma(\phi^\sigma)$ of the total Josephson current for a given edge, $\sigma$,
at a given magnetic field $B\propto\gamma$ determine the $Q$-factor, $Q^\sigma(\gamma,T)$. The latter serves as a figure of merit quantifying the SDE as a function of the out-of-plane field and temperature, 
\begin{equation}\label{eq:QfactorDef}
Q^\sigma(\gamma,T)=\frac{|I^{\sigma +}|-|I^{\sigma -}|}{I_0}\,.
\end{equation}
Combining the low-temperature limit of Eq.~(\ref{eq:JosephsonCurrentABS}) with Eq.~(\ref{eq:JosephsonCurrentContinuumApprox}), yields,
\begin{widetext}
\begin{equation}\label{eq:QT0}
Q^\sigma(\gamma,T\to0)=\sigma
\biggl\{
\Bigl(\sqrt{1-\gamma^2}-1+\frac{4\gamma}{\pi}\Bigr)\Theta(1-\gamma)+\Bigl[\frac{4}{\pi}\left(\gamma-\sqrt{\gamma^2-1}\right)+\frac{2}{\pi}\arctan\left(\sqrt{\gamma^2-1}\right)-1\Bigr]\Theta(\gamma-1)
\biggr\}.
\end{equation}
\end{widetext}

Figure \ref{fig:JosephsonCurrentTop} illustrates the behavior of the Josephson current on the top edge. In particular, the maximal and minimal Josephson currents, $I^{t +}$ and $I^{t -}$, are displayed in Fig.~\ref{fig:JosephsonCurrentTop}(a) as functions of $\gamma\propto B$, while different contributions to $I^{t \pm}$, carried by the ABS and continuum states, are plotted in Figs.~\ref{fig:JosephsonCurrentTop}(b)~and~(c).
As already observed in Fig.~3, the Josephson current gets suppressed with increasing $\gamma$, and thus also $I^{\sigma \pm}$. Consequently, the $Q$-factor is expected to approach zero when $\gamma$ increases.

\begin{figure}[t]
\centering
\includegraphics*[width=8.5cm]{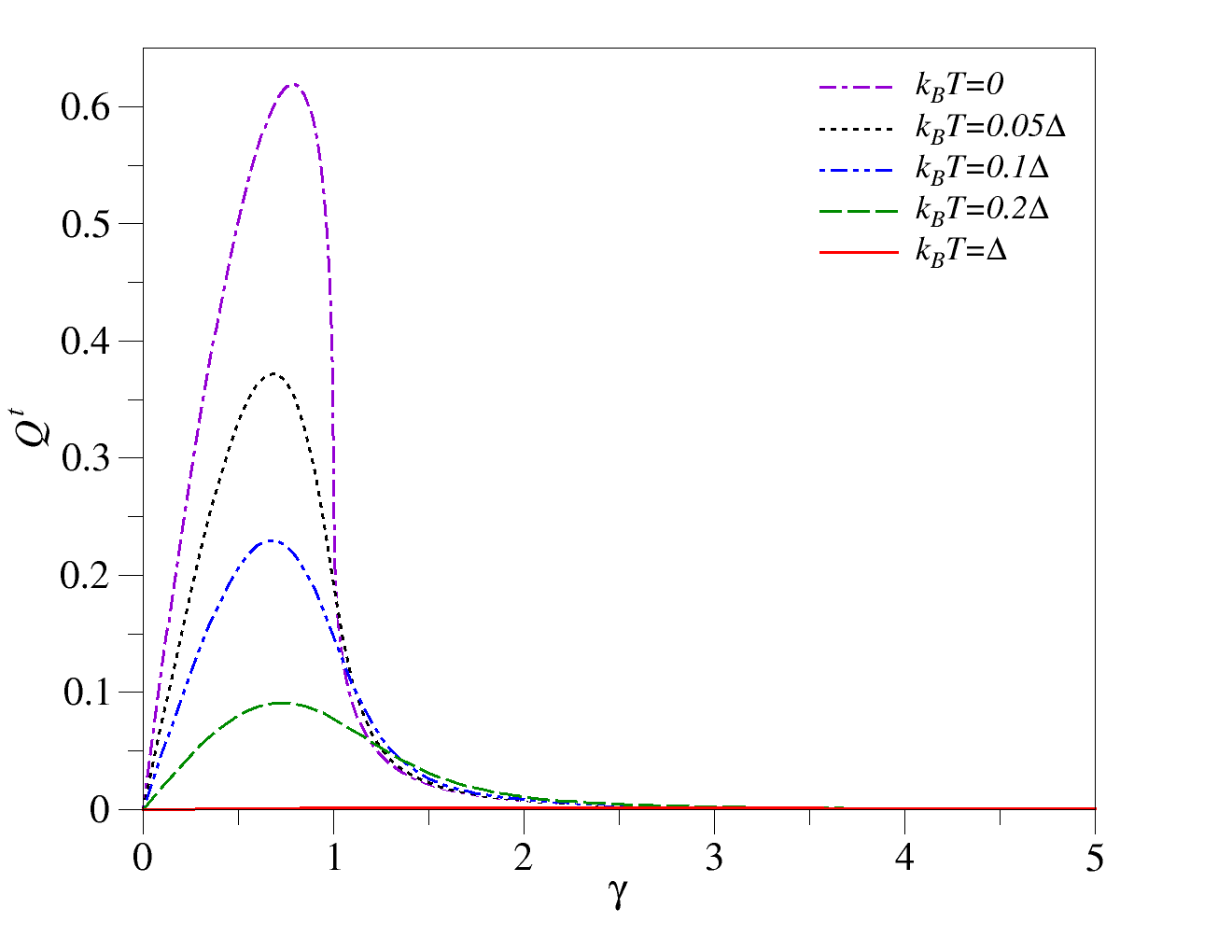}
\caption{Diode effect $Q$-factor, Eq.~(\ref{eq:QfactorDef}), for the top edge of a QSHI-based JJ as a function of $\gamma\propto B$ displayed for different temperatures. 
Here $Q^t(T\to0)$ is computed by means of Eq.~(\ref{eq:QT0}).}\label{fig:QFactorTop}
\end{figure}

This is borne out by Fig.~\ref{fig:QFactorTop}, which shows the $Q$-factor of the top edge for different temperatures. 
According to Eq.~(\ref{eq:QT0}), the $Q$-factor at $T=0$ exhibits a pronounced maximum $Q^t_{\text{max}}=\sqrt{16/\pi^2+1}-1\approx0.618993$ at $\gamma_{\text{max}}=1/\sqrt{1+\pi^2/16}\approx0.786439$. The fact that both $Q^t_{\text{max}}$ and $\gamma_{\text{max}}$ reduce to universal values, independent of junction parameters, is quite remarkable.\footnote{For the bottom edge, there is a minimum at $\gamma=1/\sqrt{1+\pi^2/16}$ with a value of $Q^b=-\sqrt{16/\pi^2+1}+1$.} 
As $T$ increases, the $Q$-factor decreases until the pronounced peak observed at low temperatures is no longer discernible [see, for example, the solid line corresponding to $k_{\text{B}}T=\Delta$ in Fig.~\ref{fig:QFactorTop}].

It is clear from Eq.~(\ref{eq:QT0}) that the net $Q$-factor of the narrow topological JJs, consisting of top and bottom edges, $Q_{\text{tot}}=Q^t+Q^b$, vanishes because $Q^t=-Q^b$ (see also Appendix~\ref{app:JJTopBottom}). Hence, to realize a finite SDE in QSHI-based junctions, a setup with disparate QSHI edges is required. This can be achieved, for example, by using the setup depicted in Fig.~\ref{fig:Setup}(a), where the superconductor proximitizes only one edge (the top edge in this case), or by designing the heterojunction in a way that the corresponding Fermi velocities $v_F^t$ and $v_F^b$ become substantially different.

\section{Parity Conserving Superconducting Diode Effect}\label{Sec:Parity}

Having discussed the SDE in a situation without constraints on the fermion parity, we now discuss a scenario in which the fermionic ground-state parity is conserved. To keep track of that, all underlying quantities are 
indexed by the subscript $p$. Correspondingly, the parity-conserved free energy---up to phase-independent contributions---reads,\cite{Ioselevich2011:PRL,Beenakker2013:PRL,Scharf2021:PRR}
\begin{widetext}
\small{
\begin{equation}\label{eq:FreeEnergyParity}
F^\sigma_p(\phi^\sigma,T)=F^\sigma(\phi^\sigma,T)-
k_{\text{B}}T\ln
\biggl\{
\frac{1}{2}
\biggl[1+pP_\sigma(\phi^\sigma)
\tanh\Bigl|\frac{\epsilon^\sigma_\uparrow(\phi^\sigma)}{2k_{\text{B}}T}\Bigr|
\exp\Bigl[
J_S(T)+\int\limits_0^\infty\d\epsilon\rho^\sigma_\mathrm{tot}(\epsilon,\phi^\sigma)\ln\Bigl[\tanh\Bigl(\frac{\epsilon}{2k_{\text{B}}T}\Bigr)\Bigr]
\Bigr]
\biggr]
\biggr\},
\end{equation}}
\end{widetext}
where the equilibrium free energy $F^\sigma(\phi^\sigma,T)$ is given by Eq.~(\ref{eq:FreeEnergy}) and the 
parity factor equals,
\begin{equation}\label{eq:DefParity}
P_\sigma(\phi^\sigma)=\mathrm{sgn}
\Bigl[
\cos\frac{\phi^\sigma}{2}-\sigma\gamma\mathrm{sgn}\Bigl(\sin\frac{\phi^\sigma}{2}\Bigr)
\Bigr].
\end{equation}
The above form of $P_\sigma(\phi^\sigma)$ implies a convention according to which the fermionic parity $p=+1$ corresponds to the lower, and $p=-1$ to the upper spectral branches in Fig.~\ref{fig:ABS}, where lower and upper refer to energies near $\phi=0$. 

In contrast to the Josephson current $I^\sigma(\phi^\sigma,T)$ derived from the equilibrium free energy $F^\sigma(\phi^\sigma,T)$, the parity conserving Josephson current $I^\sigma_p(\phi^\sigma,T)$ derived from $F_p^\sigma(\phi^\sigma,T)$ via Eq.~(\ref{eq:JosephsonCurrentDef}), 
involves the total quasi-particle DOS, 
$\rho^\sigma_\mathrm{tot}(\epsilon,\phi^\sigma)=\rho^\sigma(\epsilon,\phi^\sigma)+\overline{\rho}_0(\epsilon)$, which in addition to the term $\rho^\sigma(\epsilon,\phi^\sigma)$ given by Eq.~(\ref{eq:DOS}) contains the phase-independent contribution,
\begin{equation}\label{eq:DOS0}
\overline{\rho}_0(\epsilon)=\frac{2}{\pi E_S}\sum_{s=\uparrow/\downarrow}\frac{|\epsilon_s|\Theta\left(\epsilon_s^2-\Delta^2\right)}{\sqrt{\epsilon_s^2-\Delta^2}},
\end{equation}
where the energy scale, $E_S=\hbar v_F/L_S$, is related to the total length $L_S$ of the superconducting QSHI edge. In a similar way, the $\phi$-independent DOS of the superconducting electrodes on top of the QSHI contribute with \footnote{Assuming a quasi-particle density of states $\rho_S(\epsilon)=2/(\pi E_S)|\epsilon|\Theta\left(\epsilon^2-\Delta^2\right)/\sqrt{\epsilon^2-\Delta^2}$ from the superconducting electrodes, one can compute its contribution to Eq.~(\ref{eq:FreeEnergyParity}) via $J_S(T)=\int\limits_0^\infty\d\epsilon\rho_S(\epsilon)\ln\Bigl[\tanh\Bigl(\frac{\epsilon}{2k_{\text{B}}T}\Bigr)\Bigr]$, which yields Eq.~(\ref{eq:Js}), as detailed in Refs.~\onlinecite{Beenakker2013:PRL,Scharf2021:PRR}.}
\begin{equation}\label{eq:Js}
J_S(T)=-\frac{2}{\pi k_{\text{B}}T E_S}\int\limits_\Delta^\infty\d\epsilon\;\frac{\sqrt{\epsilon^2-\Delta^2}}{\sinh\left(\epsilon/k_{\text{B}}T\right)}.
\end{equation}

\begin{figure}[t]
\centering
\includegraphics*[width=8.5cm]{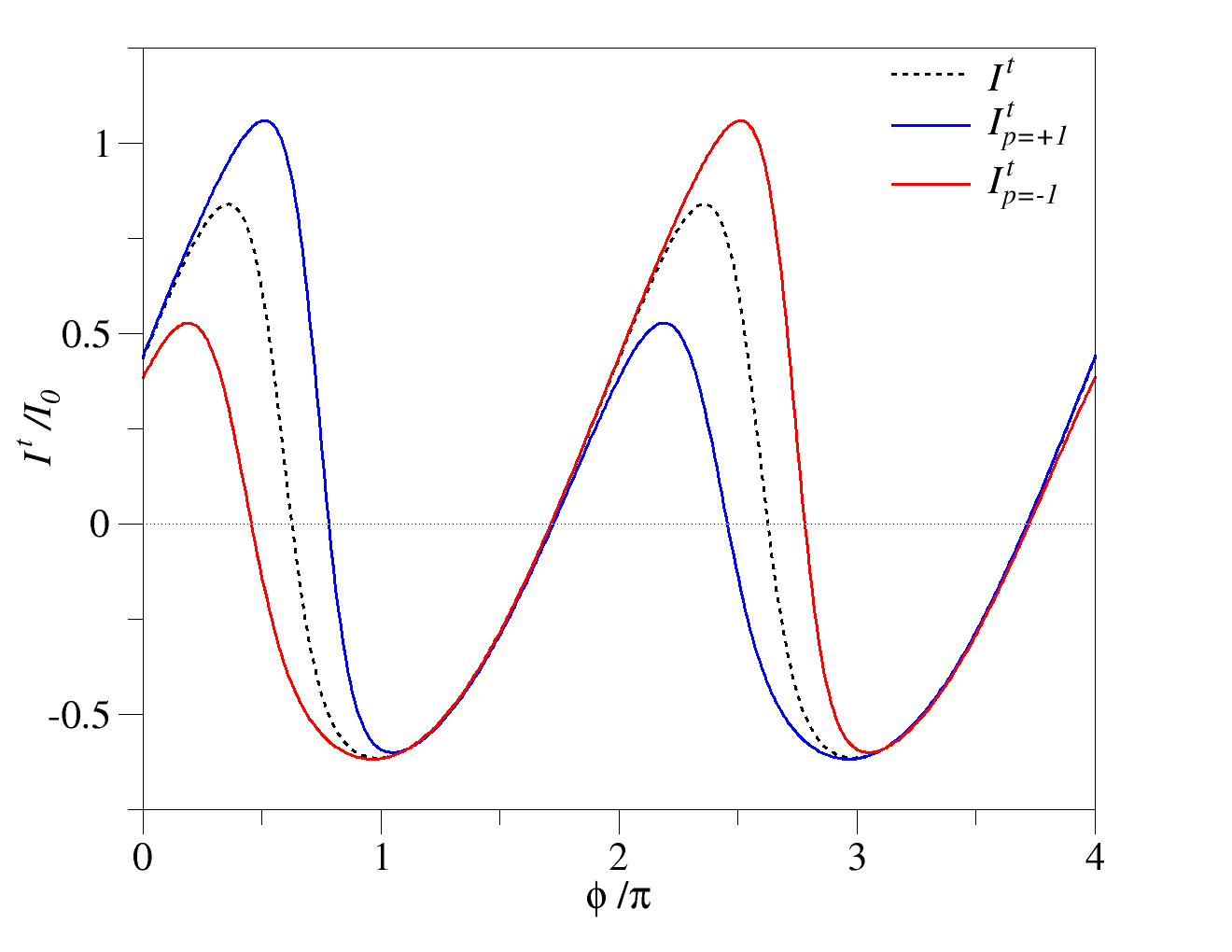}
\caption{Josephson current carried by the top edge of a QSHI-based JJ plotted as a function of the phase difference $\phi$ with (solid lines) and without (dashed line) parity constraints: $I^t$ stands for parity unconstrained $2\pi$ periodic current, Secs.~\ref{Sec:Model}-\ref{Sec:Qfactor}, while $I^t_{p=\pm1}$ denotes parity conserved $4\pi$ periodic currents with $p=\pm1$. Here $\gamma=0.6$, $k_{\text{B}}T=0.1\Delta$, $E_S=0.05\Delta$, $L_N\Delta/\hbar v_F=0.1$ and the currents are measured in units of $I_0=e\Delta/(2\hbar)$.}\label{fig:JosephsonCurrentParity}
\end{figure}

Although the underlying expressions get more involved when compared to Secs.~\ref{Sec:FreeEnergyCurrent}-\ref{Sec:Qfactor}, still some approximate analytical results can be obtained in the limiting case $T\rightarrow 0$ and $\gamma\ll 1$. However, one needs to pay an attention in which order are the corresponding mathematical operations taken into action: first goes an integration in Eq.~(\ref{eq:FreeEnergyParity}), then derivative with respect to $\phi$ and finally the limit $T\rightarrow 0$. Proceeding in this way, the current for a fixed fermionic parity $p=\pm1$ reduces to
\begin{equation}\label{eq:JosephsonCurrentParityT0}
I^\sigma_{p}(\phi^\sigma,T\to0)=\sigma I_0\left(p\sin\frac{\phi^\sigma}{2}+\frac{2\gamma}{\pi}\right),\quad\gamma\ll1
\end{equation}
with the corresponding $Q$-factor [see Eq.~(\ref{eq:QfactorDef})]
\begin{equation}\label{eq:QfactorParityT0}
Q^\sigma_{p}(\gamma,T\to0)=\sigma\frac{4\gamma}{\pi},\quad\gamma\ll1.
\end{equation}
While Eqs.~(\ref{eq:JosephsonCurrentParityT0}) and (\ref{eq:QfactorParityT0}) already provide some useful guidance, we proceed fully numerically for a more detailed analysis.

Figure~\ref{fig:JosephsonCurrentParity} shows the parity conserving Josephson currents, $I^t_{p=\pm 1}(\phi^\sigma,T)$, at the top edge, for both fermionic parities $p=\pm1$. In contrast to the parity unconstrained Josephson current, $I^t(\phi^\sigma,T)$, which exhibits $2\pi$ periodicity, its parity-conserved counterpart, $c$, becomes $4\pi$ periodic.\cite{Ioselevich2011:PRL,Beenakker2013:PRL,Scharf2021:PRR} Intriguingly, Fig.~\ref{fig:JosephsonCurrentParity} illustrates that the current-phase relations of the two different parities $p=\pm1$ are only shifted along the $\phi$ axis with respect to each other, implying the $Q$ factor to be independent of the value of $p$, as also shown by Eq.~(\ref{eq:QfactorParityT0}). Moreover, one can discern from Fig.~\ref{fig:JosephsonCurrentParity} that for $\gamma<1$ the asymmetry between the magnitudes of the maximal and minimal values of the Josephson current increases when the fermionic ground-state parity is kept conserved.
\begin{figure}[t]
\centering
\includegraphics*[width=8.5cm]{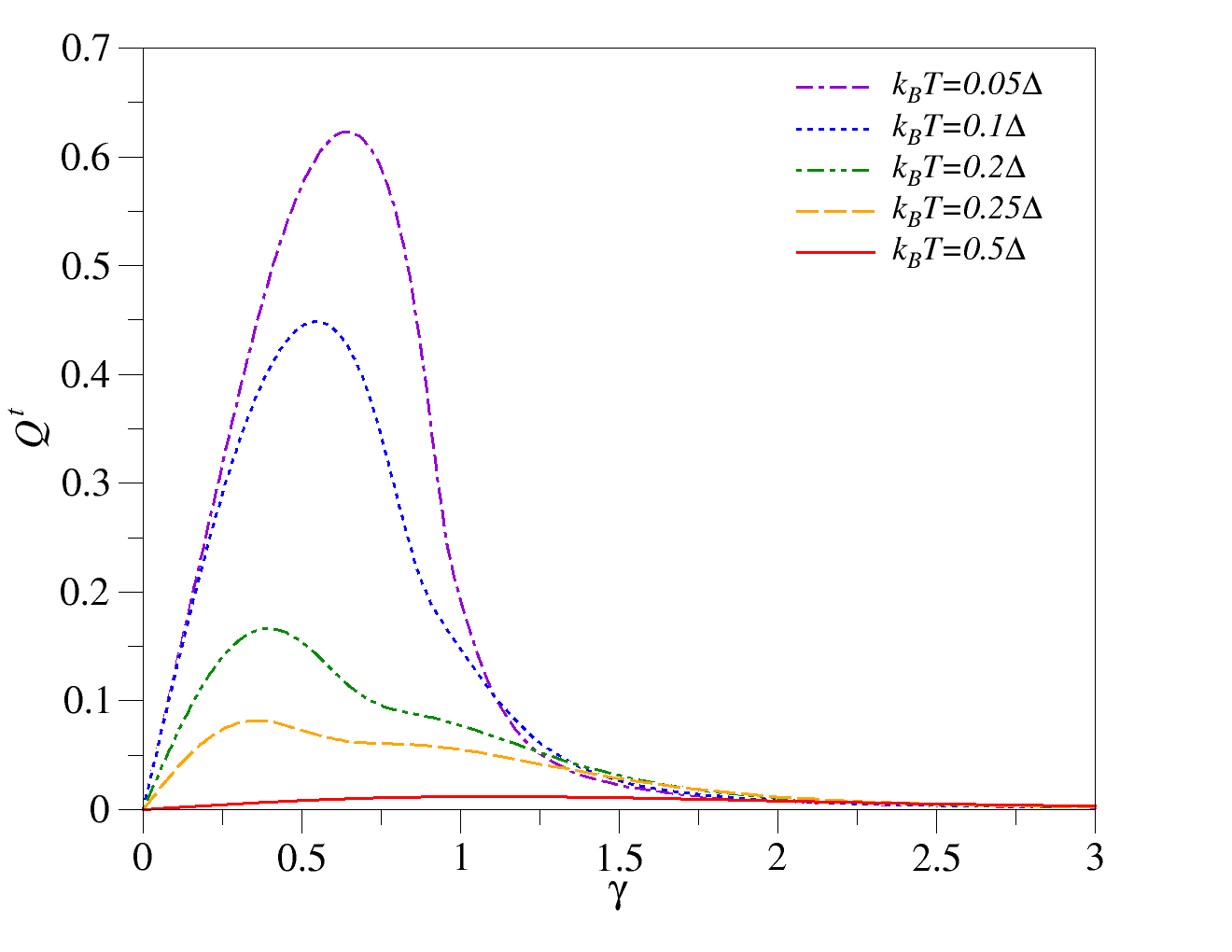}
\caption{Diode effect $Q$-factor, Eq.~(\ref{eq:QfactorDef}), for the top edge of a QSHI-based JJ with the parity $p=+1$ as a function of $\gamma\propto B$ and different temperatures. Here $E_S=0.05\Delta$. Note that the $Q$-factor for $p=-1$ is the same as that for $p=+1$.}\label{fig:QFactorTopParity}
\end{figure}
This, in turn, implies that for $\gamma < 1$ parity conservation enhances the SDE, as evidenced by Fig.~\ref{fig:QFactorTopParity}, where the magnetic-field dependence of the $Q$-factor [see Eq.~(\ref{eq:QfactorDef})] is shown for $p=+1$ and various temperatures. Indeed, by comparing Figs.~\ref{fig:QFactorTop} and \ref{fig:QFactorTopParity} we can observe that, at a given temperature, the parity-conserved $Q$-factor is generally larger than its parity-unconstrained counterpart, as long as the magnetic field keeps the global superconducting 
gap open (that is, when $\gamma<1$). The parity-enhanced $Q$-factor is reminiscent of the enhanced parity-protected SDE predicted in semiconductor-based Majorana wires.\cite{Legg2023:PRB}
However, when $\gamma\geq1$, the additional contributions in Eq.~(\ref{eq:FreeEnergyParity}) tend to zero and $F_p^\sigma(\phi^\sigma,T)$ tends to the equilibrium free energy $F^\sigma(\phi^\sigma,T)$. 
Therefore, the $Q$-factors for the case with and without fermionic parity conservation approach each other when $\gamma\geq 1$. For low temperatures, $Q$ increases linearly (see Fig.~\ref{fig:QFactorTopParity}), which is perfectly described by Eq.~(\ref{eq:QfactorParityT0}).

While we predict that for magnetic fields such that $\gamma<1$, topological JJs with conserved fermionic parity exhibit larger $Q$-factors than their parity unconstrained counterparts, their experimental realization becomes more challenging, as will be discussed in the following section.

\section{Experimental Realizations}\label{Sec:ExpReal}

Turning to potential experimental realizations displaying the SDE, we first consider the situation, where the fermionic ground-state parity does not play a role. As discussed above, if both edges of the QSHI-based JJ are equivalent [e.g., as illustrated in Fig.~\ref{fig:Setup}(b)], their contributions to the SDE cancel each other out. One way to overcome this difficulty is to design the system so that the two edges of the JJ become nonequivalent. For example, by predominantly transporting Cooper pairs along one edge, as shown in Fig.~\ref{fig:Setup}(a). In this configuration, only the top edge constitutes a short JJ, while the bottom edge forms a long JJ that carries less current due to a diminished contribution from the ABS. Consequently, the Q-factor is dominated by that of the top edge, making the single-edge physics discussed in Sec.~\ref{Sec:Qfactor}experimentally relevant.

Another way to make the edges nonequivalent is by endowing them with different Fermi velocities. For instance, in ${\rm Hg}_{1-x}{\rm Cd}_x{\rm Te}/{\rm HgTe}/{\rm Hg}_{1-x}{\rm Cd}_x{\rm Te}$-based quantum wells realizing a QSHI, the bandgap and Fermi velocity depend on (i) the thickness of the HgTe spacer,\cite{BHZ2006:S,Koenig2007:S} and (ii) the stoichiometric ratio of the Cd compound.\cite{Sengupta2013:JAP,Topalovic2020:JAP} Therefore, manufacturing HgTe-based quantum wells with different thicknesses and/or different Cd concentrations at opposite edges will make the edges nonequivalent, resulting in an observable non-zero SDE. For such HgTe-based junctions, thin-film aluminium has been used successfully to induce superconductivity in the presence of magnetic fields of more than 1 T in the normal region.\cite{Hart2017:NP,Ren2019:N} We expect much smaller magnetic fields, well below the critical magnetic field of the parent superconductor, are needed to observe a sizable SDE in such a system, however.\footnote{Assuming, for example, the system from Ref.~\onlinecite{Ren2019:N} with an induced superconducting gap of $\Delta\approx64$ $\mu$eV, a Fermi velocity of $v_F=5\times10^5$  m/s and a width of the junction of $\mathcal{W}_S=1$ $\mu$m, we obtain $\gamma\approx3.9$ for $B=1$ mT, showing that no excessively large magnetic fields are necessary to observe a sizeable SDE.}

While we anticipate a finite SDE for QSHI-junctions with a single edge or non-equivalent edges in a situation without parity constraints, preserving the fermionic parity of the ground state in experimental setups poses additional challenges, requiring conducting experiments on timescales shorter than the quasi-particle poisoning rate.\cite{Lutchyn2010:PRL,Chiu2019:PRB} For the topological JJs based on ${\rm Hg}_{1-x}{\rm Cd}_x{\rm Te}/{\rm HgTe}/{\rm Hg}_{1-x}{\rm Cd}_x{\rm Te}$ such timescales become of the order of 1\,$\mu$s.\cite{Rainis2012:PRB,Virtanen2013:PRB,Frombach2020:PRB} Consequently, experiments aiming to measure the parity-conserving SDE must operate within a sub-$\mu$s range. Nevertheless, with modern qubit and SQUID technologies enabling controlled modulation of $\phi$ on timescales of 1\,ns or shorter~\cite{Mueck2010:SST}, the observation of a parity-conserving SDE in narrow topological JJs, despite being challenging, should be experimentally feasible.

\section{Conclusions}

In this work we have studied the SDE in narrow topological QSHI-based JJs triggered by an out-of-plane magnetic field. In general, the realization of the SDE in QSHI-based JJs requires the transport of Cooper pairs through nonequivalent edge channels at the opposite ends of the junction. We investigated two different parity regimes---the conventional, so-called parity-unconstrained regime and a novel one, where the fermionic parity of the ground state is preserved. Our findings demonstrate that QSHI-based JJs can be used as versatile experimental platforms showcasing the SDE.
Furthermore, our calculations predict an increase in the Q-factor and, consequently, the diode efficiency as the temperature decreases. Interestingly, in the parity-unconstrained low-temperature regime, the maximum diode efficiency exhibits a universal character, with the maximum $Q$-factor value being independent of the system and Hamiltonian parameters. This remarkable behavior appears to be a direct consequence of the topological nature of edge-state charge transport in the QSHI regime.

Complementary to detailed numerical simulations, we also provide valuable analytical results that can be readily applied to understanding, interpreting, and fitting experimental data.


\begin{acknowledgments}
D.K.~acknowledges partial support from the project 
IM-2021-26 (SUPERSPIN) funded by the Slovak Academy of Sciences via the programme IMPULZ 2021, and
VEGA Grant No.~2/0156/22---QuaSiModo. A.M.A.~acknowledges support from ONR Grant No. N000141712793.
\end{acknowledgments}

\appendix
\section{Alternative definition of $Q$-factor}\label{app:AltQDefinition}

\begin{figure}[t]
\centering
\includegraphics*[width=8.5cm]{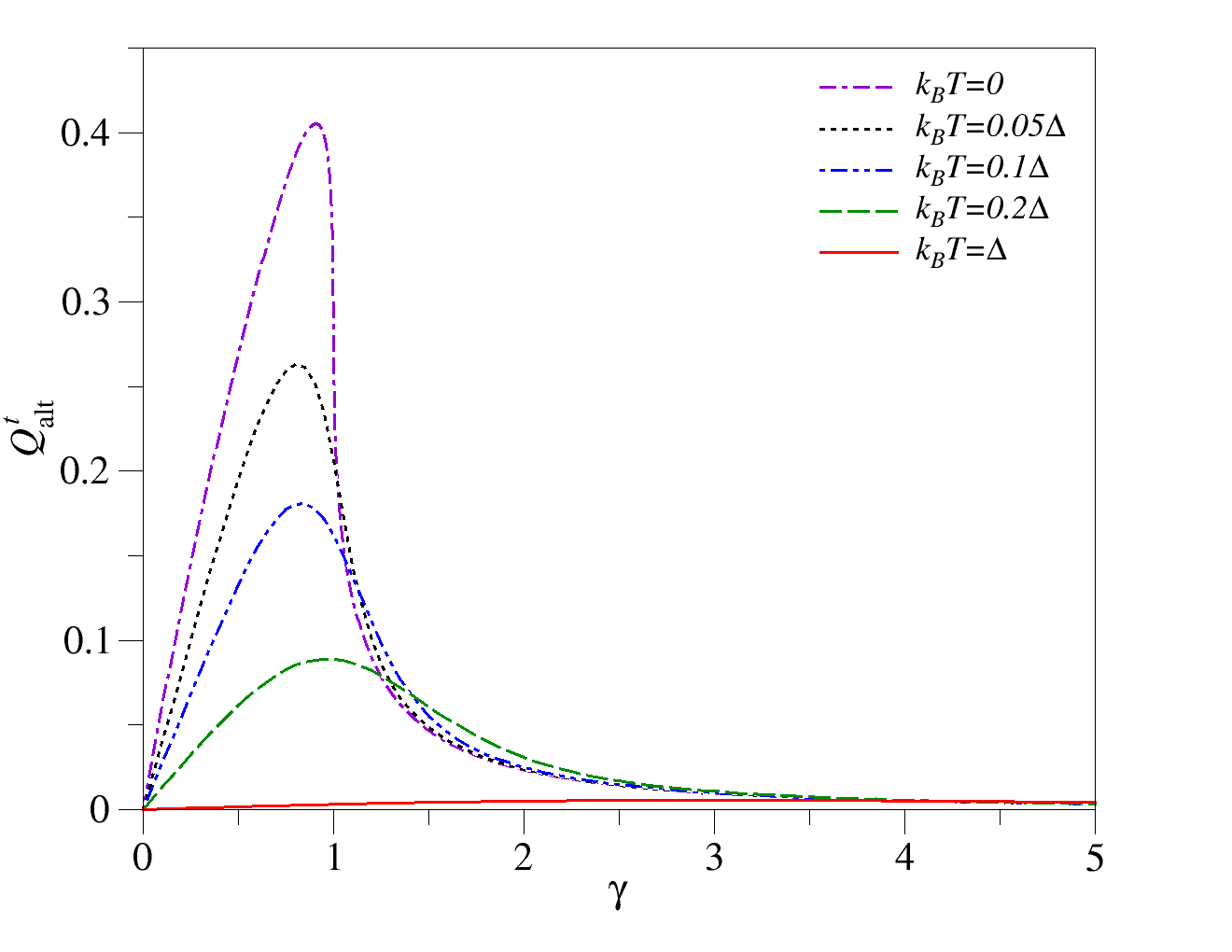}
\caption{$Q$-factor of the top edge, $Q^t_{\text{alt}}$, see Eq.~(\ref{eq:QfactorAltDef}), as a function of $\gamma=v_Fp_S/(2\Delta)\propto B$ for different temperatures. $Q(T\to 0)$ is computed from 
Eq.~(\ref{eq:QfactorAltT0}).}\label{fig:QFactorAlt}
\end{figure}

In addition to the definition of the $Q$-factor as given in Eq.~(\ref{eq:QfactorDef}), one can alternatively define the $Q$-factor as
\begin{equation}\label{eq:QfactorAltDef}
Q_{\text{alt}}^\sigma(\gamma,T)=\frac{|I^{\sigma +}|-|I^{\sigma -}|}{|I^{\sigma +}|+|I^{\sigma -}|},
\end{equation}
where $I^{\sigma +}$ and $I^{\sigma -}$ are the global maxima and minima of the Josephson current, respectively. The dependence of $Q_{\text{alt}}^t(\gamma,T)$ on $\gamma$ is shown 
in Fig.~\ref{fig:QFactorAlt} for different values of the temperature. In the zero-temperature limit the $Q$-factor defined by Eq.~(\ref{eq:QfactorAltDef}) can be written as,
\begin{widetext}
\begin{equation}\label{eq:QfactorAltT0}
Q_{\text{alt}}^\sigma(\gamma,T\to0)=\sigma\left\{\frac{\sqrt{1-\gamma^2}-1+4\gamma/\pi}{\sqrt{1-\gamma^2}+1}\Theta(1-\gamma)+\left[\frac{2\left(\gamma-\sqrt{\gamma^2-1}\right)}{\pi/2-\arctan\left(\sqrt{\gamma^2-1}\right)}-1\right]\Theta(\gamma-1)\right\}.
\end{equation}
\end{widetext}

\section{Josephson current at the top and bottom edges}\label{app:JJTopBottom}

\begin{figure}[t]
\centering
\includegraphics*[width=8.5cm]{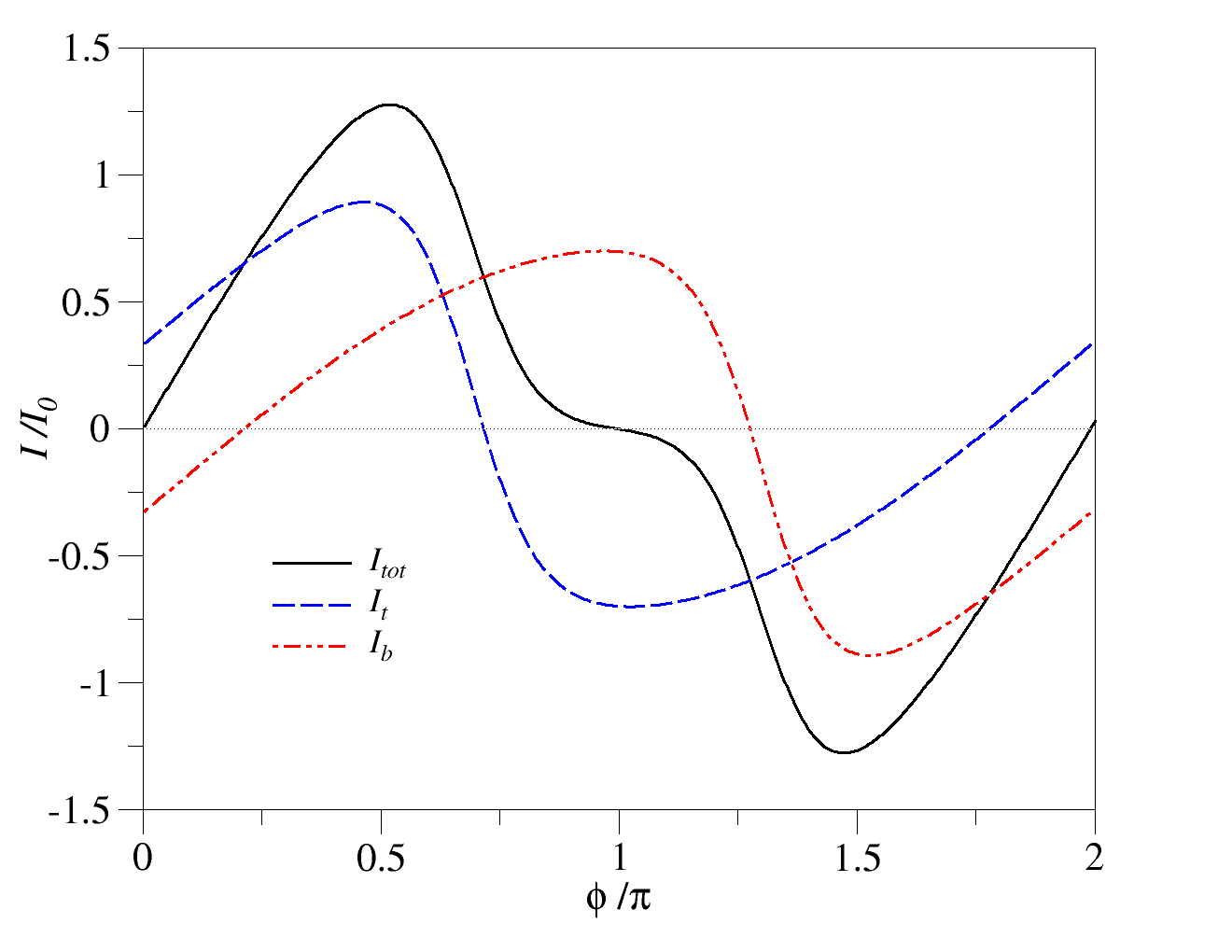}
\caption{Total Josephson current (black) of a QSHI-based JJ and the individual contributions from the top (blue) and bottom (red) edges as a function of the phase difference $\phi$. Here $v_Fp_S=0.9\Delta$ and $L_N\Delta/\hbar v_F=0.1$.}\label{fig:JosephsonCurrentTopBottom}
\end{figure}

As discussed in the main text, if the edge states at the opposite ends (say, top and bottom) of the QSHI-based JJ are equivalent, the SDE contribution at the top and bottom edges cancel each other out, leading to a vanishing $Q$-factor. This is illustrated in Fig.~\ref{fig:JosephsonCurrentTopBottom}, where the total Josephson current, $I_\text{tot}(\phi, T)=I^t(\phi, T)+I^b(\phi, T)$, as well as the individual contributions from the top and bottom edges are shown. When the edges are equivalent, their associated currents $I^t(\phi, T)$ and $I^b(\phi, T)$ obey the symmetry relation, $\max[I_\text{tot}(\phi, T)]=|\min[I_\text{tot}(\phi, T)]|$, which in turn implies that $Q_\text{tot}=0$.

\bibliography{BibTopInsAndTopSup}

\begin{thebibliography}{76}%
\makeatletter
\providecommand \@ifxundefined [1]{%
 \@ifx{#1\undefined}
}%
\providecommand \@ifnum [1]{%
 \ifnum #1\expandafter \@firstoftwo
 \else \expandafter \@secondoftwo
 \fi
}%
\providecommand \@ifx [1]{%
 \ifx #1\expandafter \@firstoftwo
 \else \expandafter \@secondoftwo
 \fi
}%
\providecommand \natexlab [1]{#1}%
\providecommand \enquote  [1]{``#1''}%
\providecommand \bibnamefont  [1]{#1}%
\providecommand \bibfnamefont [1]{#1}%
\providecommand \citenamefont [1]{#1}%
\providecommand \href@noop [0]{\@secondoftwo}%
\providecommand \href [0]{\begingroup \@sanitize@url \@href}%
\providecommand \@href[1]{\@@startlink{#1}\@@href}%
\providecommand \@@href[1]{\endgroup#1\@@endlink}%
\providecommand \@sanitize@url [0]{\catcode `\\12\catcode `\$12\catcode
  `\&12\catcode `\#12\catcode `\^12\catcode `\_12\catcode `\%12\relax}%
\providecommand \@@startlink[1]{}%
\providecommand \@@endlink[0]{}%
\providecommand \url  [0]{\begingroup\@sanitize@url \@url }%
\providecommand \@url [1]{\endgroup\@href {#1}{\urlprefix }}%
\providecommand \urlprefix  [0]{URL }%
\providecommand \Eprint [0]{\href }%
\providecommand \doibase [0]{https://doi.org/}%
\providecommand \selectlanguage [0]{\@gobble}%
\providecommand \bibinfo  [0]{\@secondoftwo}%
\providecommand \bibfield  [0]{\@secondoftwo}%
\providecommand \translation [1]{[#1]}%
\providecommand \BibitemOpen [0]{}%
\providecommand \bibitemStop [0]{}%
\providecommand \bibitemNoStop [0]{.\EOS\space}%
\providecommand \EOS [0]{\spacefactor3000\relax}%
\providecommand \BibitemShut  [1]{\csname bibitem#1\endcsname}%
\let\auto@bib@innerbib\@empty
\bibitem [{\citenamefont {Nadeem}\ \emph {et~al.}(2023)\citenamefont {Nadeem},
  \citenamefont {Fuhrer},\ and\ \citenamefont {Wang}}]{Nadeem_NatRevPhys2023}%
  \BibitemOpen
  \bibfield  {author} {\bibinfo {author} {\bibfnamefont {M.}~\bibnamefont
  {Nadeem}}, \bibinfo {author} {\bibfnamefont {M.~S.}\ \bibnamefont {Fuhrer}},\
  and\ \bibinfo {author} {\bibfnamefont {X.}~\bibnamefont {Wang}},\ }\bibfield
  {title} {\bibinfo {title} {{The superconducting diode effect}},\ }\href
  {https://doi.org/10.1038/s42254-023-00632-w} {\bibfield  {journal} {\bibinfo
  {journal} {Nature Reviews Physics}\ }\textbf {\bibinfo {volume} {5}},\
  \bibinfo {pages} {558} (\bibinfo {year} {2023})}\BibitemShut {NoStop}%
\bibitem [{\citenamefont {Sivakov}\ \emph {et~al.}(2018)\citenamefont
  {Sivakov}, \citenamefont {Turutanov}, \citenamefont {Kolinko},\ and\
  \citenamefont {Pokhila}}]{Sivakov2018:LowTempPhys-Exp}%
  \BibitemOpen
  \bibfield  {author} {\bibinfo {author} {\bibfnamefont {A.~G.}\ \bibnamefont
  {Sivakov}}, \bibinfo {author} {\bibfnamefont {O.~G.}\ \bibnamefont
  {Turutanov}}, \bibinfo {author} {\bibfnamefont {A.~E.}\ \bibnamefont
  {Kolinko}},\ and\ \bibinfo {author} {\bibfnamefont {A.~S.}\ \bibnamefont
  {Pokhila}},\ }\bibfield  {title} {\bibinfo {title} {{Spatial characterization
  of the edge barrier in wide superconducting films}},\ }\href
  {https://doi.org/10.1063/1.5024540} {\bibfield  {journal} {\bibinfo
  {journal} {Low Temperature Physics}\ }\textbf {\bibinfo {volume} {44}},\
  \bibinfo {pages} {226} (\bibinfo {year} {2018})}\BibitemShut {NoStop}%
\bibitem [{\citenamefont {Vodolazov}\ \emph {et~al.}(2018)\citenamefont
  {Vodolazov}, \citenamefont {Aladyshkin}, \citenamefont {Pestov},
  \citenamefont {Vdovichev}, \citenamefont {Ustavshikov}, \citenamefont
  {Levichev}, \citenamefont {Putilov}, \citenamefont {Yunin}, \citenamefont
  {El'kina}, \citenamefont {Bukharov},\ and\ \citenamefont
  {Klushin}}]{Vodolazov2018-SuperScinTech}%
  \BibitemOpen
  \bibfield  {author} {\bibinfo {author} {\bibfnamefont {D.~Y.}\ \bibnamefont
  {Vodolazov}}, \bibinfo {author} {\bibfnamefont {A.~Y.}\ \bibnamefont
  {Aladyshkin}}, \bibinfo {author} {\bibfnamefont {E.~E.}\ \bibnamefont
  {Pestov}}, \bibinfo {author} {\bibfnamefont {S.~N.}\ \bibnamefont
  {Vdovichev}}, \bibinfo {author} {\bibfnamefont {S.~S.}\ \bibnamefont
  {Ustavshikov}}, \bibinfo {author} {\bibfnamefont {M.~Y.}\ \bibnamefont
  {Levichev}}, \bibinfo {author} {\bibfnamefont {A.~V.}\ \bibnamefont
  {Putilov}}, \bibinfo {author} {\bibfnamefont {P.~A.}\ \bibnamefont {Yunin}},
  \bibinfo {author} {\bibfnamefont {A.~I.}\ \bibnamefont {El'kina}}, \bibinfo
  {author} {\bibfnamefont {N.~N.}\ \bibnamefont {Bukharov}},\ and\ \bibinfo
  {author} {\bibfnamefont {A.~M.}\ \bibnamefont {Klushin}},\ }\bibfield
  {title} {\bibinfo {title} {{Peculiar superconducting properties of a thin
  film superconductor{\textendash}normal metal bilayer with large ratio of
  resistivities}},\ }\href {https://doi.org/10.1088/1361-6668/aada2e}
  {\bibfield  {journal} {\bibinfo  {journal} {Supercond. Sci. Technol.}\
  }\textbf {\bibinfo {volume} {31}},\ \bibinfo {pages} {115004} (\bibinfo
  {year} {2018})}\BibitemShut {NoStop}%
\bibitem [{\citenamefont {Ando}\ \emph {et~al.}(2020)\citenamefont {Ando},
  \citenamefont {Miyasaka}, \citenamefont {Li}, \citenamefont {Ishizuka},
  \citenamefont {Arakawa}, \citenamefont {Shiota}, \citenamefont {Moriyama},
  \citenamefont {Yanase},\ and\ \citenamefont {Ono}}]{Ando2020}%
  \BibitemOpen
  \bibfield  {author} {\bibinfo {author} {\bibfnamefont {F.}~\bibnamefont
  {Ando}}, \bibinfo {author} {\bibfnamefont {Y.}~\bibnamefont {Miyasaka}},
  \bibinfo {author} {\bibfnamefont {T.}~\bibnamefont {Li}}, \bibinfo {author}
  {\bibfnamefont {J.}~\bibnamefont {Ishizuka}}, \bibinfo {author}
  {\bibfnamefont {T.}~\bibnamefont {Arakawa}}, \bibinfo {author} {\bibfnamefont
  {Y.}~\bibnamefont {Shiota}}, \bibinfo {author} {\bibfnamefont
  {T.}~\bibnamefont {Moriyama}}, \bibinfo {author} {\bibfnamefont
  {Y.}~\bibnamefont {Yanase}},\ and\ \bibinfo {author} {\bibfnamefont
  {T.}~\bibnamefont {Ono}},\ }\bibfield  {title} {\bibinfo {title}
  {{Observation of superconducting diode effect}},\ }\href
  {https://doi.org/10.1038/s41586-020-2590-4} {\bibfield  {journal} {\bibinfo
  {journal} {Nature}\ }\textbf {\bibinfo {volume} {584}},\ \bibinfo {pages}
  {373} (\bibinfo {year} {2020})}\BibitemShut {NoStop}%
\bibitem [{\citenamefont {Hou}\ \emph {et~al.}(2023)\citenamefont {Hou},
  \citenamefont {Nichele}, \citenamefont {Chi}, \citenamefont {Lodesani},
  \citenamefont {Wu}, \citenamefont {Ritter}, \citenamefont {Haxell},
  \citenamefont {Davydova}, \citenamefont {Ili\ifmmode~\acute{c}\else
  \'{c}\fi{}}, \citenamefont {Glezakou-Elbert}, \citenamefont {Varambally},
  \citenamefont {Bergeret}, \citenamefont {Kamra}, \citenamefont {Fu},
  \citenamefont {Lee},\ and\ \citenamefont {Moodera}}]{Hou2023:PRL}%
  \BibitemOpen
  \bibfield  {author} {\bibinfo {author} {\bibfnamefont {Y.}~\bibnamefont
  {Hou}}, \bibinfo {author} {\bibfnamefont {F.}~\bibnamefont {Nichele}},
  \bibinfo {author} {\bibfnamefont {H.}~\bibnamefont {Chi}}, \bibinfo {author}
  {\bibfnamefont {A.}~\bibnamefont {Lodesani}}, \bibinfo {author}
  {\bibfnamefont {Y.}~\bibnamefont {Wu}}, \bibinfo {author} {\bibfnamefont
  {M.~F.}\ \bibnamefont {Ritter}}, \bibinfo {author} {\bibfnamefont {D.~Z.}\
  \bibnamefont {Haxell}}, \bibinfo {author} {\bibfnamefont {M.}~\bibnamefont
  {Davydova}}, \bibinfo {author} {\bibfnamefont {S.}~\bibnamefont
  {Ili\ifmmode~\acute{c}\else \'{c}\fi{}}}, \bibinfo {author} {\bibfnamefont
  {O.}~\bibnamefont {Glezakou-Elbert}}, \bibinfo {author} {\bibfnamefont
  {A.}~\bibnamefont {Varambally}}, \bibinfo {author} {\bibfnamefont {F.~S.}\
  \bibnamefont {Bergeret}}, \bibinfo {author} {\bibfnamefont {A.}~\bibnamefont
  {Kamra}}, \bibinfo {author} {\bibfnamefont {L.}~\bibnamefont {Fu}}, \bibinfo
  {author} {\bibfnamefont {P.~A.}\ \bibnamefont {Lee}},\ and\ \bibinfo {author}
  {\bibfnamefont {J.~S.}\ \bibnamefont {Moodera}},\ }\bibfield  {title}
  {\bibinfo {title} {{Ubiquitous Superconducting Diode Effect in Superconductor
  Thin Films}},\ }\href {https://doi.org/10.1103/PhysRevLett.131.027001}
  {\bibfield  {journal} {\bibinfo  {journal} {Phys. Rev. Lett.}\ }\textbf
  {\bibinfo {volume} {131}},\ \bibinfo {pages} {027001} (\bibinfo {year}
  {2023})}\BibitemShut {NoStop}%
\bibitem [{\citenamefont {Baumgartner}\ \emph
  {et~al.}(2022{\natexlab{a}})\citenamefont {Baumgartner}, \citenamefont
  {Fuchs}, \citenamefont {Costa}, \citenamefont {Reinhardt}, \citenamefont
  {Gronin}, \citenamefont {Gardner}, \citenamefont {Lindemann}, \citenamefont
  {Manfra}, \citenamefont {Faria~Junior}, \citenamefont {Kochan}, \citenamefont
  {Fabian}, \citenamefont {Paradiso},\ and\ \citenamefont
  {Strunk}}]{Baumgartner2022}%
  \BibitemOpen
  \bibfield  {author} {\bibinfo {author} {\bibfnamefont {C.}~\bibnamefont
  {Baumgartner}}, \bibinfo {author} {\bibfnamefont {L.}~\bibnamefont {Fuchs}},
  \bibinfo {author} {\bibfnamefont {A.}~\bibnamefont {Costa}}, \bibinfo
  {author} {\bibfnamefont {S.}~\bibnamefont {Reinhardt}}, \bibinfo {author}
  {\bibfnamefont {S.}~\bibnamefont {Gronin}}, \bibinfo {author} {\bibfnamefont
  {G.~C.}\ \bibnamefont {Gardner}}, \bibinfo {author} {\bibfnamefont
  {T.}~\bibnamefont {Lindemann}}, \bibinfo {author} {\bibfnamefont {M.~J.}\
  \bibnamefont {Manfra}}, \bibinfo {author} {\bibfnamefont {P.~E.}\
  \bibnamefont {Faria~Junior}}, \bibinfo {author} {\bibfnamefont
  {D.}~\bibnamefont {Kochan}}, \bibinfo {author} {\bibfnamefont
  {J.}~\bibnamefont {Fabian}}, \bibinfo {author} {\bibfnamefont
  {N.}~\bibnamefont {Paradiso}},\ and\ \bibinfo {author} {\bibfnamefont
  {C.}~\bibnamefont {Strunk}},\ }\bibfield  {title} {\bibinfo {title}
  {{Supercurrent rectification and magnetochiral effects in symmetric Josephson
  junctions}},\ }\href {https://doi.org/10.1038/s41565-021-01009-9} {\bibfield
  {journal} {\bibinfo  {journal} {Nature Nanotechnology}\ }\textbf {\bibinfo
  {volume} {17}},\ \bibinfo {pages} {39} (\bibinfo {year}
  {2022}{\natexlab{a}})}\BibitemShut {NoStop}%
\bibitem [{\citenamefont {Baumgartner}\ \emph
  {et~al.}(2022{\natexlab{b}})\citenamefont {Baumgartner}, \citenamefont
  {Fuchs}, \citenamefont {Costa}, \citenamefont {Pic{\'{o}}-Cort{\'{e}}s},
  \citenamefont {Reinhardt}, \citenamefont {Gronin}, \citenamefont {Gardner},
  \citenamefont {Lindemann}, \citenamefont {Manfra}, \citenamefont {Junior},
  \citenamefont {Kochan}, \citenamefont {Fabian}, \citenamefont {Paradiso},\
  and\ \citenamefont {Strunk}}]{BaumgartnerSI2022}%
  \BibitemOpen
  \bibfield  {author} {\bibinfo {author} {\bibfnamefont {C.}~\bibnamefont
  {Baumgartner}}, \bibinfo {author} {\bibfnamefont {L.}~\bibnamefont {Fuchs}},
  \bibinfo {author} {\bibfnamefont {A.}~\bibnamefont {Costa}}, \bibinfo
  {author} {\bibfnamefont {J.}~\bibnamefont {Pic{\'{o}}-Cort{\'{e}}s}},
  \bibinfo {author} {\bibfnamefont {S.}~\bibnamefont {Reinhardt}}, \bibinfo
  {author} {\bibfnamefont {S.}~\bibnamefont {Gronin}}, \bibinfo {author}
  {\bibfnamefont {G.~C.}\ \bibnamefont {Gardner}}, \bibinfo {author}
  {\bibfnamefont {T.}~\bibnamefont {Lindemann}}, \bibinfo {author}
  {\bibfnamefont {M.~J.}\ \bibnamefont {Manfra}}, \bibinfo {author}
  {\bibfnamefont {P.~E.~F.}\ \bibnamefont {Junior}}, \bibinfo {author}
  {\bibfnamefont {D.}~\bibnamefont {Kochan}}, \bibinfo {author} {\bibfnamefont
  {J.}~\bibnamefont {Fabian}}, \bibinfo {author} {\bibfnamefont
  {N.}~\bibnamefont {Paradiso}},\ and\ \bibinfo {author} {\bibfnamefont
  {C.}~\bibnamefont {Strunk}},\ }\bibfield  {title} {\bibinfo {title} {{Effect
  of Rashba and Dresselhaus spin{\textendash}orbit coupling on supercurrent
  rectification and magnetochiral anisotropy of ballistic Josephson
  junctions}},\ }\href {https://doi.org/10.1088/1361-648x/ac4d5e} {\bibfield
  {journal} {\bibinfo  {journal} {Journal of Physics: Condensed Matter}\
  }\textbf {\bibinfo {volume} {34}},\ \bibinfo {pages} {154005} (\bibinfo
  {year} {2022}{\natexlab{b}})}\BibitemShut {NoStop}%
\bibitem [{\citenamefont {Turini}\ \emph {et~al.}(2022)\citenamefont {Turini},
  \citenamefont {Salimian}, \citenamefont {Carrega}, \citenamefont {Iorio},
  \citenamefont {Strambini}, \citenamefont {Giazotto}, \citenamefont {Zannier},
  \citenamefont {Sorba},\ and\ \citenamefont {Heun}}]{Turini2022}%
  \BibitemOpen
  \bibfield  {author} {\bibinfo {author} {\bibfnamefont {B.}~\bibnamefont
  {Turini}}, \bibinfo {author} {\bibfnamefont {S.}~\bibnamefont {Salimian}},
  \bibinfo {author} {\bibfnamefont {M.}~\bibnamefont {Carrega}}, \bibinfo
  {author} {\bibfnamefont {A.}~\bibnamefont {Iorio}}, \bibinfo {author}
  {\bibfnamefont {E.}~\bibnamefont {Strambini}}, \bibinfo {author}
  {\bibfnamefont {F.}~\bibnamefont {Giazotto}}, \bibinfo {author}
  {\bibfnamefont {V.}~\bibnamefont {Zannier}}, \bibinfo {author} {\bibfnamefont
  {L.}~\bibnamefont {Sorba}},\ and\ \bibinfo {author} {\bibfnamefont
  {S.}~\bibnamefont {Heun}},\ }\bibfield  {title} {\bibinfo {title} {Josephson
  diode effect in high-mobility insb nanoflags},\ }\href
  {https://doi.org/10.1021/acs.nanolett.2c02899} {\bibfield  {journal}
  {\bibinfo  {journal} {Nano Letters}\ }\textbf {\bibinfo {volume} {22}},\
  \bibinfo {pages} {8502} (\bibinfo {year} {2022})}\BibitemShut {NoStop}%
\bibitem [{\citenamefont {Costa}\ \emph
  {et~al.}(2023{\natexlab{a}})\citenamefont {Costa}, \citenamefont
  {Baumgartner}, \citenamefont {Reinhardt}, \citenamefont {Berger},
  \citenamefont {Gronin}, \citenamefont {Gardner}, \citenamefont {Lindemann},
  \citenamefont {Manfra}, \citenamefont {Fabian}, \citenamefont {Kochan},
  \citenamefont {Paradiso},\ and\ \citenamefont
  {Strunk}}]{Costa2023:NatureNanotechnology}%
  \BibitemOpen
  \bibfield  {author} {\bibinfo {author} {\bibfnamefont {A.}~\bibnamefont
  {Costa}}, \bibinfo {author} {\bibfnamefont {C.}~\bibnamefont {Baumgartner}},
  \bibinfo {author} {\bibfnamefont {S.}~\bibnamefont {Reinhardt}}, \bibinfo
  {author} {\bibfnamefont {J.}~\bibnamefont {Berger}}, \bibinfo {author}
  {\bibfnamefont {S.}~\bibnamefont {Gronin}}, \bibinfo {author} {\bibfnamefont
  {G.~C.}\ \bibnamefont {Gardner}}, \bibinfo {author} {\bibfnamefont
  {T.}~\bibnamefont {Lindemann}}, \bibinfo {author} {\bibfnamefont {M.~J.}\
  \bibnamefont {Manfra}}, \bibinfo {author} {\bibfnamefont {J.}~\bibnamefont
  {Fabian}}, \bibinfo {author} {\bibfnamefont {D.}~\bibnamefont {Kochan}},
  \bibinfo {author} {\bibfnamefont {N.}~\bibnamefont {Paradiso}},\ and\
  \bibinfo {author} {\bibfnamefont {C.}~\bibnamefont {Strunk}},\ }\bibfield
  {title} {\bibinfo {title} {Sign reversal of the josephson inductance
  magnetochiral anisotropy and $0$--$\pi$-like transitions in supercurrent
  diodes},\ }\href {https://doi.org/10.1038/s41565-023-01451-x} {\bibfield
  {journal} {\bibinfo  {journal} {Nature Nanotechnology}\ }\textbf {\bibinfo
  {volume} {18}},\ \bibinfo {pages} {1266} (\bibinfo {year}
  {2023}{\natexlab{a}})}\BibitemShut {NoStop}%
\bibitem [{\citenamefont {Lotfizadeh}\ \emph {et~al.}(2024)\citenamefont
  {Lotfizadeh}, \citenamefont {Schiela}, \citenamefont {Pekerten},
  \citenamefont {Yu}, \citenamefont {Elfeky}, \citenamefont {Strickland},
  \citenamefont {Matos-Abiague},\ and\ \citenamefont
  {Shabani}}]{Lotfizadeh:CP2024}%
  \BibitemOpen
  \bibfield  {author} {\bibinfo {author} {\bibfnamefont {N.}~\bibnamefont
  {Lotfizadeh}}, \bibinfo {author} {\bibfnamefont {W.~F.}\ \bibnamefont
  {Schiela}}, \bibinfo {author} {\bibfnamefont {B.}~\bibnamefont {Pekerten}},
  \bibinfo {author} {\bibfnamefont {P.}~\bibnamefont {Yu}}, \bibinfo {author}
  {\bibfnamefont {B.~H.}\ \bibnamefont {Elfeky}}, \bibinfo {author}
  {\bibfnamefont {W.~M.}\ \bibnamefont {Strickland}}, \bibinfo {author}
  {\bibfnamefont {A.}~\bibnamefont {Matos-Abiague}},\ and\ \bibinfo {author}
  {\bibfnamefont {J.}~\bibnamefont {Shabani}},\ }\bibfield  {title} {\bibinfo
  {title} {Superconducting diode effect sign change in epitaxial al-inas
  josephson junctions},\ }\href@noop {} {\bibfield  {journal} {\bibinfo
  {journal} {Commun. Phys.}\ }\textbf {\bibinfo {volume} {7}},\ \bibinfo
  {pages} {120} (\bibinfo {year} {2024})}\BibitemShut {NoStop}%
\bibitem [{\citenamefont {Reinhardt}\ \emph {et~al.}(2024)\citenamefont
  {Reinhardt}, \citenamefont {Ascherl}, \citenamefont {Costa}, \citenamefont
  {Berger}, \citenamefont {Gronin}, \citenamefont {Gardner}, \citenamefont
  {Lindemann}, \citenamefont {Manfra}, \citenamefont {Fabian}, \citenamefont
  {Kochan}, \citenamefont {Strunk},\ and\ \citenamefont
  {Paradiso}}]{Reinhardt2024:NatureCommm}%
  \BibitemOpen
  \bibfield  {author} {\bibinfo {author} {\bibfnamefont {S.}~\bibnamefont
  {Reinhardt}}, \bibinfo {author} {\bibfnamefont {T.}~\bibnamefont {Ascherl}},
  \bibinfo {author} {\bibfnamefont {A.}~\bibnamefont {Costa}}, \bibinfo
  {author} {\bibfnamefont {J.}~\bibnamefont {Berger}}, \bibinfo {author}
  {\bibfnamefont {S.}~\bibnamefont {Gronin}}, \bibinfo {author} {\bibfnamefont
  {G.~C.}\ \bibnamefont {Gardner}}, \bibinfo {author} {\bibfnamefont
  {T.}~\bibnamefont {Lindemann}}, \bibinfo {author} {\bibfnamefont {M.~J.}\
  \bibnamefont {Manfra}}, \bibinfo {author} {\bibfnamefont {J.}~\bibnamefont
  {Fabian}}, \bibinfo {author} {\bibfnamefont {D.}~\bibnamefont {Kochan}},
  \bibinfo {author} {\bibfnamefont {C.}~\bibnamefont {Strunk}},\ and\ \bibinfo
  {author} {\bibfnamefont {N.}~\bibnamefont {Paradiso}},\ }\bibfield  {title}
  {\bibinfo {title} {{Link between supercurrent diode and anomalous Josephson
  effect revealed by gate-controlled interferometry}},\ }\href
  {https://doi.org/10.1038/s41467-024-48741-z} {\bibfield  {journal} {\bibinfo
  {journal} {Nature Communications}\ }\textbf {\bibinfo {volume} {15}},\
  \bibinfo {pages} {4413} (\bibinfo {year} {2024})}\BibitemShut {NoStop}%
\bibitem [{\citenamefont {Pal}\ \emph {et~al.}(2022)\citenamefont {Pal},
  \citenamefont {Chakraborty}, \citenamefont {Sivakumar}, \citenamefont
  {Davydova}, \citenamefont {Gopi}, \citenamefont {Pandeya}, \citenamefont
  {Krieger}, \citenamefont {Zhang}, \citenamefont {Date}, \citenamefont {Ju},
  \citenamefont {Yuan}, \citenamefont {Schr{\"o}ter}, \citenamefont {Fu},\ and\
  \citenamefont {Parkin}}]{Pal2022}%
  \BibitemOpen
  \bibfield  {author} {\bibinfo {author} {\bibfnamefont {B.}~\bibnamefont
  {Pal}}, \bibinfo {author} {\bibfnamefont {A.}~\bibnamefont {Chakraborty}},
  \bibinfo {author} {\bibfnamefont {P.~K.}\ \bibnamefont {Sivakumar}}, \bibinfo
  {author} {\bibfnamefont {M.}~\bibnamefont {Davydova}}, \bibinfo {author}
  {\bibfnamefont {A.~K.}\ \bibnamefont {Gopi}}, \bibinfo {author}
  {\bibfnamefont {A.~K.}\ \bibnamefont {Pandeya}}, \bibinfo {author}
  {\bibfnamefont {J.~A.}\ \bibnamefont {Krieger}}, \bibinfo {author}
  {\bibfnamefont {Y.}~\bibnamefont {Zhang}}, \bibinfo {author} {\bibfnamefont
  {M.}~\bibnamefont {Date}}, \bibinfo {author} {\bibfnamefont {S.}~\bibnamefont
  {Ju}}, \bibinfo {author} {\bibfnamefont {N.}~\bibnamefont {Yuan}}, \bibinfo
  {author} {\bibfnamefont {N.~B.~M.}\ \bibnamefont {Schr{\"o}ter}}, \bibinfo
  {author} {\bibfnamefont {L.}~\bibnamefont {Fu}},\ and\ \bibinfo {author}
  {\bibfnamefont {S.~S.~P.}\ \bibnamefont {Parkin}},\ }\bibfield  {title}
  {\bibinfo {title} {{Josephson diode effect from Cooper pair momentum in a
  topological semimetal}},\ }\href {https://doi.org/10.1038/s41567-022-01699-5}
  {\bibfield  {journal} {\bibinfo  {journal} {Nature Physics}\ } (\bibinfo
  {year} {2022})}\BibitemShut {NoStop}%
\bibitem [{\citenamefont {Jeon}\ \emph {et~al.}(2022)\citenamefont {Jeon},
  \citenamefont {Kim}, \citenamefont {Yoon}, \citenamefont {Jeon},
  \citenamefont {Han}, \citenamefont {Cottet}, \citenamefont {Kontos},\ and\
  \citenamefont {Parkin}}]{Jeon2022}%
  \BibitemOpen
  \bibfield  {author} {\bibinfo {author} {\bibfnamefont {K.-R.}\ \bibnamefont
  {Jeon}}, \bibinfo {author} {\bibfnamefont {J.-K.}\ \bibnamefont {Kim}},
  \bibinfo {author} {\bibfnamefont {J.}~\bibnamefont {Yoon}}, \bibinfo {author}
  {\bibfnamefont {J.-C.}\ \bibnamefont {Jeon}}, \bibinfo {author}
  {\bibfnamefont {H.}~\bibnamefont {Han}}, \bibinfo {author} {\bibfnamefont
  {A.}~\bibnamefont {Cottet}}, \bibinfo {author} {\bibfnamefont
  {T.}~\bibnamefont {Kontos}},\ and\ \bibinfo {author} {\bibfnamefont
  {S.~S.~P.}\ \bibnamefont {Parkin}},\ }\bibfield  {title} {\bibinfo {title}
  {{Zero-field polarity-reversible Josephson supercurrent diodes enabled by a
  proximity-magnetized Pt barrier}},\ }\href
  {https://doi.org/10.1038/s41563-022-01300-7} {\bibfield  {journal} {\bibinfo
  {journal} {Nature Materials}\ }\textbf {\bibinfo {volume} {21}},\ \bibinfo
  {pages} {1008} (\bibinfo {year} {2022})}\BibitemShut {NoStop}%
\bibitem [{\citenamefont {Wakatsuki}\ \emph {et~al.}(2017)\citenamefont
  {Wakatsuki}, \citenamefont {Saito}, \citenamefont {Hoshino}, \citenamefont
  {Itahashi}, \citenamefont {Ideue}, \citenamefont {Ezawa}, \citenamefont
  {Iwasa},\ and\ \citenamefont {Nagaosa}}]{Wakatsuki2017}%
  \BibitemOpen
  \bibfield  {author} {\bibinfo {author} {\bibfnamefont {R.}~\bibnamefont
  {Wakatsuki}}, \bibinfo {author} {\bibfnamefont {Y.}~\bibnamefont {Saito}},
  \bibinfo {author} {\bibfnamefont {S.}~\bibnamefont {Hoshino}}, \bibinfo
  {author} {\bibfnamefont {Y.~M.}\ \bibnamefont {Itahashi}}, \bibinfo {author}
  {\bibfnamefont {T.}~\bibnamefont {Ideue}}, \bibinfo {author} {\bibfnamefont
  {M.}~\bibnamefont {Ezawa}}, \bibinfo {author} {\bibfnamefont
  {Y.}~\bibnamefont {Iwasa}},\ and\ \bibinfo {author} {\bibfnamefont
  {N.}~\bibnamefont {Nagaosa}},\ }\bibfield  {title} {\bibinfo {title}
  {{Nonreciprocal charge transport in noncentrosymmetric superconductors}},\
  }\href {https://doi.org/10.1126/sciadv.1602390} {\bibfield  {journal}
  {\bibinfo  {journal} {Science Advances}\ }\textbf {\bibinfo {volume} {3}},\
  \bibinfo {pages} {e1602390} (\bibinfo {year} {2017})}\BibitemShut {NoStop}%
\bibitem [{\citenamefont {Wu}\ \emph {et~al.}(2022)\citenamefont {Wu},
  \citenamefont {Wang}, \citenamefont {Xu}, \citenamefont {Sivakumar},
  \citenamefont {Pasco}, \citenamefont {Filippozzi}, \citenamefont {Parkin},
  \citenamefont {Zeng}, \citenamefont {McQueen},\ and\ \citenamefont
  {Ali}}]{Wu2022}%
  \BibitemOpen
  \bibfield  {author} {\bibinfo {author} {\bibfnamefont {H.}~\bibnamefont
  {Wu}}, \bibinfo {author} {\bibfnamefont {Y.}~\bibnamefont {Wang}}, \bibinfo
  {author} {\bibfnamefont {Y.}~\bibnamefont {Xu}}, \bibinfo {author}
  {\bibfnamefont {P.~K.}\ \bibnamefont {Sivakumar}}, \bibinfo {author}
  {\bibfnamefont {C.}~\bibnamefont {Pasco}}, \bibinfo {author} {\bibfnamefont
  {U.}~\bibnamefont {Filippozzi}}, \bibinfo {author} {\bibfnamefont {S.~S.~P.}\
  \bibnamefont {Parkin}}, \bibinfo {author} {\bibfnamefont {Y.-J.}\
  \bibnamefont {Zeng}}, \bibinfo {author} {\bibfnamefont {T.}~\bibnamefont
  {McQueen}},\ and\ \bibinfo {author} {\bibfnamefont {M.~N.}\ \bibnamefont
  {Ali}},\ }\bibfield  {title} {\bibinfo {title} {{The field-free Josephson
  diode in a van der Waals heterostructure}},\ }\href
  {https://doi.org/10.1038/s41586-022-04504-8} {\bibfield  {journal} {\bibinfo
  {journal} {Nature}\ }\textbf {\bibinfo {volume} {604}},\ \bibinfo {pages}
  {653} (\bibinfo {year} {2022})}\BibitemShut {NoStop}%
\bibitem [{\citenamefont {Bauriedl}\ \emph {et~al.}(2022)\citenamefont
  {Bauriedl}, \citenamefont {B{\"a}uml}, \citenamefont {Fuchs}, \citenamefont
  {Baumgartner}, \citenamefont {Paulik}, \citenamefont {Bauer}, \citenamefont
  {Lin}, \citenamefont {Lupton}, \citenamefont {Taniguchi}, \citenamefont
  {Watanabe}, \citenamefont {Strunk},\ and\ \citenamefont
  {Paradiso}}]{Bauriedl2022}%
  \BibitemOpen
  \bibfield  {author} {\bibinfo {author} {\bibfnamefont {L.}~\bibnamefont
  {Bauriedl}}, \bibinfo {author} {\bibfnamefont {C.}~\bibnamefont {B{\"a}uml}},
  \bibinfo {author} {\bibfnamefont {L.}~\bibnamefont {Fuchs}}, \bibinfo
  {author} {\bibfnamefont {C.}~\bibnamefont {Baumgartner}}, \bibinfo {author}
  {\bibfnamefont {N.}~\bibnamefont {Paulik}}, \bibinfo {author} {\bibfnamefont
  {J.~M.}\ \bibnamefont {Bauer}}, \bibinfo {author} {\bibfnamefont {K.-Q.}\
  \bibnamefont {Lin}}, \bibinfo {author} {\bibfnamefont {J.~M.}\ \bibnamefont
  {Lupton}}, \bibinfo {author} {\bibfnamefont {T.}~\bibnamefont {Taniguchi}},
  \bibinfo {author} {\bibfnamefont {K.}~\bibnamefont {Watanabe}}, \bibinfo
  {author} {\bibfnamefont {C.}~\bibnamefont {Strunk}},\ and\ \bibinfo {author}
  {\bibfnamefont {N.}~\bibnamefont {Paradiso}},\ }\bibfield  {title} {\bibinfo
  {title} {{Supercurrent diode effect and magnetochiral anisotropy in few-layer
  NbSe2}},\ }\href {https://doi.org/10.1038/s41467-022-31954-5} {\bibfield
  {journal} {\bibinfo  {journal} {Nature Communications}\ }\textbf {\bibinfo
  {volume} {13}},\ \bibinfo {pages} {4266} (\bibinfo {year}
  {2022})}\BibitemShut {NoStop}%
\bibitem [{\citenamefont {D\'{i}ez-M{\'{e}}rida}\ \emph
  {et~al.}(2023)\citenamefont {D\'{i}ez-M{\'{e}}rida}, \citenamefont
  {D\'{i}ez-Carl{\'{o}}n}, \citenamefont {Yang}, \citenamefont {Xie},
  \citenamefont {Gao}, \citenamefont {Senior}, \citenamefont {Watanabe},
  \citenamefont {Taniguchi}, \citenamefont {Lu}, \citenamefont {Higginbotham},
  \citenamefont {Law},\ and\ \citenamefont {Efetov}}]{DiezMerida2023:NatComm}%
  \BibitemOpen
  \bibfield  {author} {\bibinfo {author} {\bibfnamefont {J.}~\bibnamefont
  {D\'{i}ez-M{\'{e}}rida}}, \bibinfo {author} {\bibfnamefont {A.}~\bibnamefont
  {D\'{i}ez-Carl{\'{o}}n}}, \bibinfo {author} {\bibfnamefont {S.~Y.}\
  \bibnamefont {Yang}}, \bibinfo {author} {\bibfnamefont {Y.~M.}\ \bibnamefont
  {Xie}}, \bibinfo {author} {\bibfnamefont {X.~J.}\ \bibnamefont {Gao}},
  \bibinfo {author} {\bibfnamefont {J.}~\bibnamefont {Senior}}, \bibinfo
  {author} {\bibfnamefont {K.}~\bibnamefont {Watanabe}}, \bibinfo {author}
  {\bibfnamefont {T.}~\bibnamefont {Taniguchi}}, \bibinfo {author}
  {\bibfnamefont {X.}~\bibnamefont {Lu}}, \bibinfo {author} {\bibfnamefont
  {A.~P.}\ \bibnamefont {Higginbotham}}, \bibinfo {author} {\bibfnamefont
  {K.~T.}\ \bibnamefont {Law}},\ and\ \bibinfo {author} {\bibfnamefont {D.~K.}\
  \bibnamefont {Efetov}},\ }\bibfield  {title} {\bibinfo {title}
  {{Symmetry-broken Josephson junctions and superconducting diodes in
  magic-angle twisted bilayer graphene}},\ }\href
  {https://doi.org/10.1038/s41467-023-38005-7} {\bibfield  {journal} {\bibinfo
  {journal} {Nature Communications}\ }\textbf {\bibinfo {volume} {14}},\
  \bibinfo {pages} {2396} (\bibinfo {year} {2023})}\BibitemShut {NoStop}%
\bibitem [{\citenamefont {Lin}\ \emph {et~al.}(2022)\citenamefont {Lin},
  \citenamefont {Siriviboon}, \citenamefont {Scammell}, \citenamefont {Liu},
  \citenamefont {Rhodes}, \citenamefont {Watanabe}, \citenamefont {Taniguchi},
  \citenamefont {Hone}, \citenamefont {Scheurer},\ and\ \citenamefont
  {Li}}]{Lin2022:NatPhys-experiment}%
  \BibitemOpen
  \bibfield  {author} {\bibinfo {author} {\bibfnamefont {J.-X.}\ \bibnamefont
  {Lin}}, \bibinfo {author} {\bibfnamefont {P.}~\bibnamefont {Siriviboon}},
  \bibinfo {author} {\bibfnamefont {H.~D.}\ \bibnamefont {Scammell}}, \bibinfo
  {author} {\bibfnamefont {S.}~\bibnamefont {Liu}}, \bibinfo {author}
  {\bibfnamefont {D.}~\bibnamefont {Rhodes}}, \bibinfo {author} {\bibfnamefont
  {K.}~\bibnamefont {Watanabe}}, \bibinfo {author} {\bibfnamefont
  {T.}~\bibnamefont {Taniguchi}}, \bibinfo {author} {\bibfnamefont
  {J.}~\bibnamefont {Hone}}, \bibinfo {author} {\bibfnamefont {M.~S.}\
  \bibnamefont {Scheurer}},\ and\ \bibinfo {author} {\bibfnamefont {J.~I.~A.}\
  \bibnamefont {Li}},\ }\bibfield  {title} {\bibinfo {title} {{Zero-field
  superconducting diode effect in small-twist-angle trilayer graphene}},\
  }\href {https://doi.org/10.1038/s41567-022-01700-1} {\bibfield  {journal}
  {\bibinfo  {journal} {Nature Physics}\ }\textbf {\bibinfo {volume} {18}},\
  \bibinfo {pages} {1221} (\bibinfo {year} {2022})}\BibitemShut {NoStop}%
\bibitem [{\citenamefont {Zhao}\ \emph {et~al.}(2023)\citenamefont {Zhao},
  \citenamefont {Cui}, \citenamefont {Volkov}, \citenamefont {Yoo},
  \citenamefont {Lee}, \citenamefont {Gardener}, \citenamefont {Akey},
  \citenamefont {Engelke}, \citenamefont {Ronen}, \citenamefont {Zhong},
  \citenamefont {Gu}, \citenamefont {Plugge}, \citenamefont {Tummuru},
  \citenamefont {Kim}, \citenamefont {Franz}, \citenamefont {Pixley},
  \citenamefont {Poccia},\ and\ \citenamefont {Kim}}]{Zhao2023:S}%
  \BibitemOpen
  \bibfield  {author} {\bibinfo {author} {\bibfnamefont {S.~Y.~F.}\
  \bibnamefont {Zhao}}, \bibinfo {author} {\bibfnamefont {X.}~\bibnamefont
  {Cui}}, \bibinfo {author} {\bibfnamefont {P.~A.}\ \bibnamefont {Volkov}},
  \bibinfo {author} {\bibfnamefont {H.}~\bibnamefont {Yoo}}, \bibinfo {author}
  {\bibfnamefont {S.}~\bibnamefont {Lee}}, \bibinfo {author} {\bibfnamefont
  {J.~A.}\ \bibnamefont {Gardener}}, \bibinfo {author} {\bibfnamefont {A.~J.}\
  \bibnamefont {Akey}}, \bibinfo {author} {\bibfnamefont {R.}~\bibnamefont
  {Engelke}}, \bibinfo {author} {\bibfnamefont {Y.}~\bibnamefont {Ronen}},
  \bibinfo {author} {\bibfnamefont {R.}~\bibnamefont {Zhong}}, \bibinfo
  {author} {\bibfnamefont {G.}~\bibnamefont {Gu}}, \bibinfo {author}
  {\bibfnamefont {S.}~\bibnamefont {Plugge}}, \bibinfo {author} {\bibfnamefont
  {T.}~\bibnamefont {Tummuru}}, \bibinfo {author} {\bibfnamefont
  {M.}~\bibnamefont {Kim}}, \bibinfo {author} {\bibfnamefont {M.}~\bibnamefont
  {Franz}}, \bibinfo {author} {\bibfnamefont {J.~H.}\ \bibnamefont {Pixley}},
  \bibinfo {author} {\bibfnamefont {N.}~\bibnamefont {Poccia}},\ and\ \bibinfo
  {author} {\bibfnamefont {P.}~\bibnamefont {Kim}},\ }\bibfield  {title}
  {\bibinfo {title} {Time-reversal symmetry breaking superconductivity between
  twisted cuprate superconductors},\ }\href
  {https://doi.org/10.1126/science.abl8371} {\bibfield  {journal} {\bibinfo
  {journal} {Science}\ }\textbf {\bibinfo {volume} {382}},\ \bibinfo {pages}
  {1422} (\bibinfo {year} {2023})}\BibitemShut {NoStop}%
\bibitem [{\citenamefont {Ghosh}\ \emph {et~al.}(2024)\citenamefont {Ghosh},
  \citenamefont {Patil}, \citenamefont {Basu}, \citenamefont {Kuldeep},
  \citenamefont {Dutta}, \citenamefont {Jangade}, \citenamefont {Kulkarni},
  \citenamefont {Thamizhavel}, \citenamefont {Steiner}, \citenamefont {von
  Oppen},\ and\ \citenamefont {Deshmukh}}]{Ghosh2024:NM}%
  \BibitemOpen
  \bibfield  {author} {\bibinfo {author} {\bibfnamefont {S.}~\bibnamefont
  {Ghosh}}, \bibinfo {author} {\bibfnamefont {V.}~\bibnamefont {Patil}},
  \bibinfo {author} {\bibfnamefont {A.}~\bibnamefont {Basu}}, \bibinfo {author}
  {\bibnamefont {Kuldeep}}, \bibinfo {author} {\bibfnamefont {A.}~\bibnamefont
  {Dutta}}, \bibinfo {author} {\bibfnamefont {D.~A.}\ \bibnamefont {Jangade}},
  \bibinfo {author} {\bibfnamefont {R.}~\bibnamefont {Kulkarni}}, \bibinfo
  {author} {\bibfnamefont {A.}~\bibnamefont {Thamizhavel}}, \bibinfo {author}
  {\bibfnamefont {J.~F.}\ \bibnamefont {Steiner}}, \bibinfo {author}
  {\bibfnamefont {F.}~\bibnamefont {von Oppen}},\ and\ \bibinfo {author}
  {\bibfnamefont {M.~M.}\ \bibnamefont {Deshmukh}},\ }\bibfield  {title}
  {\bibinfo {title} {High-temperature josephson diode},\ }\href
  {https://doi.org/10.1038/s41563-024-01804-4} {\bibfield  {journal} {\bibinfo
  {journal} {Nat. Mater.}\ }\textbf {\bibinfo {volume} {23}},\ \bibinfo {pages}
  {612} (\bibinfo {year} {2024})}\BibitemShut {NoStop}%
\bibitem [{\citenamefont {Pal}\ and\ \citenamefont
  {Benjamin}(2019)}]{Pal2019:EPL}%
  \BibitemOpen
  \bibfield  {author} {\bibinfo {author} {\bibfnamefont {S.}~\bibnamefont
  {Pal}}\ and\ \bibinfo {author} {\bibfnamefont {C.}~\bibnamefont {Benjamin}},\
  }\bibfield  {title} {\bibinfo {title} {{Quantized Josephson phase battery}},\
  }\href {https://doi.org/10.1209/0295-5075/126/57002} {\bibfield  {journal}
  {\bibinfo  {journal} {Europhysics Letters}\ }\textbf {\bibinfo {volume}
  {126}},\ \bibinfo {pages} {57002} (\bibinfo {year} {2019})}\BibitemShut
  {NoStop}%
\bibitem [{\citenamefont {Chen}\ \emph {et~al.}(2018)\citenamefont {Chen},
  \citenamefont {He}, \citenamefont {Ali}, \citenamefont {Lee}, \citenamefont
  {Fong},\ and\ \citenamefont {Law}}]{Chen2018:PRB}%
  \BibitemOpen
  \bibfield  {author} {\bibinfo {author} {\bibfnamefont {C.-Z.}\ \bibnamefont
  {Chen}}, \bibinfo {author} {\bibfnamefont {J.~J.}\ \bibnamefont {He}},
  \bibinfo {author} {\bibfnamefont {M.~N.}\ \bibnamefont {Ali}}, \bibinfo
  {author} {\bibfnamefont {G.-H.}\ \bibnamefont {Lee}}, \bibinfo {author}
  {\bibfnamefont {K.~C.}\ \bibnamefont {Fong}},\ and\ \bibinfo {author}
  {\bibfnamefont {K.~T.}\ \bibnamefont {Law}},\ }\bibfield  {title} {\bibinfo
  {title} {Asymmetric josephson effect in inversion symmetry breaking
  topological materials},\ }\href {https://doi.org/10.1103/PhysRevB.98.075430}
  {\bibfield  {journal} {\bibinfo  {journal} {Phys. Rev. B}\ }\textbf {\bibinfo
  {volume} {98}},\ \bibinfo {pages} {075430} (\bibinfo {year}
  {2018})}\BibitemShut {NoStop}%
\bibitem [{\citenamefont {Tanaka}\ \emph {et~al.}(2022)\citenamefont {Tanaka},
  \citenamefont {Lu},\ and\ \citenamefont {Nagaosa}}]{Tanaka2022:PRB}%
  \BibitemOpen
  \bibfield  {author} {\bibinfo {author} {\bibfnamefont {Y.}~\bibnamefont
  {Tanaka}}, \bibinfo {author} {\bibfnamefont {B.}~\bibnamefont {Lu}},\ and\
  \bibinfo {author} {\bibfnamefont {N.}~\bibnamefont {Nagaosa}},\ }\bibfield
  {title} {\bibinfo {title} {Theory of giant diode effect in $d$-wave
  superconductor junctions on the surface of a topological insulator},\ }\href
  {https://doi.org/10.1103/PhysRevB.106.214524} {\bibfield  {journal} {\bibinfo
   {journal} {Phys. Rev. B}\ }\textbf {\bibinfo {volume} {106}},\ \bibinfo
  {pages} {214524} (\bibinfo {year} {2022})}\BibitemShut {NoStop}%
\bibitem [{\citenamefont {Lu}\ \emph {et~al.}(2023)\citenamefont {Lu},
  \citenamefont {Ikegaya}, \citenamefont {Burset}, \citenamefont {Tanaka},\
  and\ \citenamefont {Nagaosa}}]{Lu2023:PRL}%
  \BibitemOpen
  \bibfield  {author} {\bibinfo {author} {\bibfnamefont {B.}~\bibnamefont
  {Lu}}, \bibinfo {author} {\bibfnamefont {S.}~\bibnamefont {Ikegaya}},
  \bibinfo {author} {\bibfnamefont {P.}~\bibnamefont {Burset}}, \bibinfo
  {author} {\bibfnamefont {Y.}~\bibnamefont {Tanaka}},\ and\ \bibinfo {author}
  {\bibfnamefont {N.}~\bibnamefont {Nagaosa}},\ }\bibfield  {title} {\bibinfo
  {title} {Tunable josephson diode effect on the surface of topological
  insulators},\ }\href {https://doi.org/10.1103/PhysRevLett.131.096001}
  {\bibfield  {journal} {\bibinfo  {journal} {Phys. Rev. Lett.}\ }\textbf
  {\bibinfo {volume} {131}},\ \bibinfo {pages} {096001} (\bibinfo {year}
  {2023})}\BibitemShut {NoStop}%
\bibitem [{\citenamefont {Fu}\ \emph {et~al.}(2024{\natexlab{a}})\citenamefont
  {Fu}, \citenamefont {Xu}, \citenamefont {Yang}, \citenamefont {Lee},
  \citenamefont {Ang},\ and\ \citenamefont {Liu}}]{Fu2024:PRApp}%
  \BibitemOpen
  \bibfield  {author} {\bibinfo {author} {\bibfnamefont {P.-H.}\ \bibnamefont
  {Fu}}, \bibinfo {author} {\bibfnamefont {Y.}~\bibnamefont {Xu}}, \bibinfo
  {author} {\bibfnamefont {S.~A.}\ \bibnamefont {Yang}}, \bibinfo {author}
  {\bibfnamefont {C.~H.}\ \bibnamefont {Lee}}, \bibinfo {author} {\bibfnamefont
  {Y.~S.}\ \bibnamefont {Ang}},\ and\ \bibinfo {author} {\bibfnamefont {J.-F.}\
  \bibnamefont {Liu}},\ }\bibfield  {title} {\bibinfo {title} {Field-effect
  josephson diode via asymmetric spin-momentum locking states},\ }\href
  {https://doi.org/10.1103/PhysRevApplied.21.054057} {\bibfield  {journal}
  {\bibinfo  {journal} {Phys. Rev. Appl.}\ }\textbf {\bibinfo {volume} {21}},\
  \bibinfo {pages} {054057} (\bibinfo {year} {2024}{\natexlab{a}})}\BibitemShut
  {NoStop}%
\bibitem [{\citenamefont {Legg}\ \emph {et~al.}(2023)\citenamefont {Legg},
  \citenamefont {Laubscher}, \citenamefont {Loss},\ and\ \citenamefont
  {Klinovaja}}]{Legg2023:PRB}%
  \BibitemOpen
  \bibfield  {author} {\bibinfo {author} {\bibfnamefont {H.~F.}\ \bibnamefont
  {Legg}}, \bibinfo {author} {\bibfnamefont {K.}~\bibnamefont {Laubscher}},
  \bibinfo {author} {\bibfnamefont {D.}~\bibnamefont {Loss}},\ and\ \bibinfo
  {author} {\bibfnamefont {J.}~\bibnamefont {Klinovaja}},\ }\bibfield  {title}
  {\bibinfo {title} {Parity-protected superconducting diode effect in
  topological josephson junctions},\ }\href
  {https://doi.org/10.1103/PhysRevB.108.214520} {\bibfield  {journal} {\bibinfo
   {journal} {Phys. Rev. B}\ }\textbf {\bibinfo {volume} {108}},\ \bibinfo
  {pages} {214520} (\bibinfo {year} {2023})}\BibitemShut {NoStop}%
\bibitem [{\citenamefont {Cayao}\ \emph {et~al.}(2024)\citenamefont {Cayao},
  \citenamefont {Nagaosa},\ and\ \citenamefont {Tanaka}}]{Cayao2024:PRB}%
  \BibitemOpen
  \bibfield  {author} {\bibinfo {author} {\bibfnamefont {J.}~\bibnamefont
  {Cayao}}, \bibinfo {author} {\bibfnamefont {N.}~\bibnamefont {Nagaosa}},\
  and\ \bibinfo {author} {\bibfnamefont {Y.}~\bibnamefont {Tanaka}},\
  }\bibfield  {title} {\bibinfo {title} {Enhancing the josephson diode effect
  with majorana bound states},\ }\href
  {https://doi.org/10.1103/PhysRevB.109.L081405} {\bibfield  {journal}
  {\bibinfo  {journal} {Phys. Rev. B}\ }\textbf {\bibinfo {volume} {109}},\
  \bibinfo {pages} {L081405} (\bibinfo {year} {2024})}\BibitemShut {NoStop}%
\bibitem [{\citenamefont {Meyer}\ and\ \citenamefont
  {Houzet}(2024)}]{Meyer2024:APL}%
  \BibitemOpen
  \bibfield  {author} {\bibinfo {author} {\bibfnamefont {J.~S.}\ \bibnamefont
  {Meyer}}\ and\ \bibinfo {author} {\bibfnamefont {M.}~\bibnamefont {Houzet}},\
  }\bibfield  {title} {\bibinfo {title} {{Josephson diode effect in a ballistic
  single-channel nanowire}},\ }\href {https://doi.org/10.1063/5.0211491}
  {\bibfield  {journal} {\bibinfo  {journal} {Appl. Phys. Lett.}\ }\textbf
  {\bibinfo {volume} {125}},\ \bibinfo {pages} {022603} (\bibinfo {year}
  {2024})}\BibitemShut {NoStop}%
\bibitem [{\citenamefont {Debnath}\ and\ \citenamefont
  {Dutta}(2024)}]{Debnath2024:PRB}%
  \BibitemOpen
  \bibfield  {author} {\bibinfo {author} {\bibfnamefont {D.}~\bibnamefont
  {Debnath}}\ and\ \bibinfo {author} {\bibfnamefont {P.}~\bibnamefont
  {Dutta}},\ }\bibfield  {title} {\bibinfo {title} {Gate-tunable josephson
  diode effect in rashba spin-orbit coupled quantum dot junctions},\ }\href
  {https://doi.org/10.1103/PhysRevB.109.174511} {\bibfield  {journal} {\bibinfo
   {journal} {Phys. Rev. B}\ }\textbf {\bibinfo {volume} {109}},\ \bibinfo
  {pages} {174511} (\bibinfo {year} {2024})}\BibitemShut {NoStop}%
\bibitem [{\citenamefont {Trahms}\ \emph {et~al.}(2023)\citenamefont {Trahms},
  \citenamefont {Melischek}, \citenamefont {Steiner}, \citenamefont {Mahendru},
  \citenamefont {Tamir}, \citenamefont {Bogdanoff}, \citenamefont {Peters},
  \citenamefont {Reecht}, \citenamefont {Winkelmann}, \citenamefont {von
  Oppen},\ and\ \citenamefont {Franke}}]{Trahms2023:Nature}%
  \BibitemOpen
  \bibfield  {author} {\bibinfo {author} {\bibfnamefont {M.}~\bibnamefont
  {Trahms}}, \bibinfo {author} {\bibfnamefont {L.}~\bibnamefont {Melischek}},
  \bibinfo {author} {\bibfnamefont {J.~F.}\ \bibnamefont {Steiner}}, \bibinfo
  {author} {\bibfnamefont {B.}~\bibnamefont {Mahendru}}, \bibinfo {author}
  {\bibfnamefont {I.}~\bibnamefont {Tamir}}, \bibinfo {author} {\bibfnamefont
  {N.}~\bibnamefont {Bogdanoff}}, \bibinfo {author} {\bibfnamefont
  {O.}~\bibnamefont {Peters}}, \bibinfo {author} {\bibfnamefont
  {G.}~\bibnamefont {Reecht}}, \bibinfo {author} {\bibfnamefont {C.~B.}\
  \bibnamefont {Winkelmann}}, \bibinfo {author} {\bibfnamefont
  {F.}~\bibnamefont {von Oppen}},\ and\ \bibinfo {author} {\bibfnamefont
  {K.~J.}\ \bibnamefont {Franke}},\ }\bibfield  {title} {\bibinfo {title}
  {{Diode effect in Josephson junctions with a single magnetic atom}},\ }\href
  {https://doi.org/10.1038/s41586-023-05743-z} {\bibfield  {journal} {\bibinfo
  {journal} {Nature}\ }\textbf {\bibinfo {volume} {615}},\ \bibinfo {pages}
  {628} (\bibinfo {year} {2023})}\BibitemShut {NoStop}%
\bibitem [{\citenamefont {Banerjee}\ and\ \citenamefont
  {Scheurer}(2024)}]{banerjee2024altermagnetic}%
  \BibitemOpen
  \bibfield  {author} {\bibinfo {author} {\bibfnamefont {S.}~\bibnamefont
  {Banerjee}}\ and\ \bibinfo {author} {\bibfnamefont {M.~S.}\ \bibnamefont
  {Scheurer}},\ }\href@noop {} {\bibinfo {title} {Altermagnetic superconducting
  diode effect}} (\bibinfo {year} {2024}),\ \Eprint
  {https://arxiv.org/abs/2402.14071} {arXiv:2402.14071 [cond-mat.supr-con]}
  \BibitemShut {NoStop}%
\bibitem [{\citenamefont {Zhang}\ \emph {et~al.}(2024)\citenamefont {Zhang},
  \citenamefont {Hu},\ and\ \citenamefont
  {Neupert}}]{Zhang-Neupert2024:NatComm}%
  \BibitemOpen
  \bibfield  {author} {\bibinfo {author} {\bibfnamefont {S.-B.}\ \bibnamefont
  {Zhang}}, \bibinfo {author} {\bibfnamefont {L.-H.}\ \bibnamefont {Hu}},\ and\
  \bibinfo {author} {\bibfnamefont {T.}~\bibnamefont {Neupert}},\ }\bibfield
  {title} {\bibinfo {title} {Finite-momentum cooper pairing in proximitized
  altermagnets},\ }\href {https://doi.org/10.1038/s41467-024-45951-3}
  {\bibfield  {journal} {\bibinfo  {journal} {Nature Communications}\ }\textbf
  {\bibinfo {volume} {15}},\ \bibinfo {pages} {1801} (\bibinfo {year}
  {2024})}\BibitemShut {NoStop}%
\bibitem [{\citenamefont {Davydova}\ \emph {et~al.}(2022)\citenamefont
  {Davydova}, \citenamefont {Prembabu},\ and\ \citenamefont
  {Fu}}]{Davydova2022}%
  \BibitemOpen
  \bibfield  {author} {\bibinfo {author} {\bibfnamefont {M.}~\bibnamefont
  {Davydova}}, \bibinfo {author} {\bibfnamefont {S.}~\bibnamefont {Prembabu}},\
  and\ \bibinfo {author} {\bibfnamefont {L.}~\bibnamefont {Fu}},\ }\bibfield
  {title} {\bibinfo {title} {Universal josephson diode effect},\ }\href
  {https://doi.org/10.1126/sciadv.abo0309} {\bibfield  {journal} {\bibinfo
  {journal} {Science Advances}\ }\textbf {\bibinfo {volume} {8}},\ \bibinfo
  {pages} {eabo0309} (\bibinfo {year} {2022})}\BibitemShut {NoStop}%
\bibitem [{\citenamefont {Fracassi}\ \emph {et~al.}(2024)\citenamefont
  {Fracassi}, \citenamefont {Traverso}, \citenamefont {Traverso~Ziani},
  \citenamefont {Carrega}, \citenamefont {Heun},\ and\ \citenamefont
  {Sassetti}}]{Fracassi2024:APL}%
  \BibitemOpen
  \bibfield  {author} {\bibinfo {author} {\bibfnamefont {S.}~\bibnamefont
  {Fracassi}}, \bibinfo {author} {\bibfnamefont {S.}~\bibnamefont {Traverso}},
  \bibinfo {author} {\bibfnamefont {N.}~\bibnamefont {Traverso~Ziani}},
  \bibinfo {author} {\bibfnamefont {M.}~\bibnamefont {Carrega}}, \bibinfo
  {author} {\bibfnamefont {S.}~\bibnamefont {Heun}},\ and\ \bibinfo {author}
  {\bibfnamefont {M.}~\bibnamefont {Sassetti}},\ }\bibfield  {title} {\bibinfo
  {title} {{Anomalous supercurrent and diode effect in locally perturbed
  topological Josephson junctions}},\ }\href
  {https://doi.org/10.1063/5.0210660} {\bibfield  {journal} {\bibinfo
  {journal} {Appl. Phys. Lett.}\ }\textbf {\bibinfo {volume} {124}},\ \bibinfo
  {pages} {242601} (\bibinfo {year} {2024})}\BibitemShut {NoStop}%
\bibitem [{\citenamefont {Coraiola}\ \emph {et~al.}(2024)\citenamefont
  {Coraiola}, \citenamefont {Svetogorov}, \citenamefont {Haxell}, \citenamefont
  {Sabonis}, \citenamefont {Hinderling}, \citenamefont {ten Kate},
  \citenamefont {Cheah}, \citenamefont {Krizek}, \citenamefont {Schott},
  \citenamefont {Wegscheider}, \citenamefont {Cuevas}, \citenamefont {Belzig},\
  and\ \citenamefont {Nichele}}]{Coraiola:ASCNano2024}%
  \BibitemOpen
  \bibfield  {author} {\bibinfo {author} {\bibfnamefont {M.}~\bibnamefont
  {Coraiola}}, \bibinfo {author} {\bibfnamefont {A.~E.}\ \bibnamefont
  {Svetogorov}}, \bibinfo {author} {\bibfnamefont {D.~Z.}\ \bibnamefont
  {Haxell}}, \bibinfo {author} {\bibfnamefont {D.}~\bibnamefont {Sabonis}},
  \bibinfo {author} {\bibfnamefont {M.}~\bibnamefont {Hinderling}}, \bibinfo
  {author} {\bibfnamefont {S.~C.}\ \bibnamefont {ten Kate}}, \bibinfo {author}
  {\bibfnamefont {E.}~\bibnamefont {Cheah}}, \bibinfo {author} {\bibfnamefont
  {F.}~\bibnamefont {Krizek}}, \bibinfo {author} {\bibfnamefont
  {R.}~\bibnamefont {Schott}}, \bibinfo {author} {\bibfnamefont
  {W.}~\bibnamefont {Wegscheider}}, \bibinfo {author} {\bibfnamefont {J.~C.}\
  \bibnamefont {Cuevas}}, \bibinfo {author} {\bibfnamefont {W.}~\bibnamefont
  {Belzig}},\ and\ \bibinfo {author} {\bibfnamefont {F.}~\bibnamefont
  {Nichele}},\ }\bibfield  {title} {\bibinfo {title} {{Flux-Tunable Josephson
  Diode Effect in a Hybrid Four-Terminal Josephson Junction}},\ }\href
  {https://doi.org/10.1021/acsnano.4c01642} {\bibfield  {journal} {\bibinfo
  {journal} {ACS Nano}\ }\textbf {\bibinfo {volume} {18}},\ \bibinfo {pages}
  {9221} (\bibinfo {year} {2024})}\BibitemShut {NoStop}%
\bibitem [{\citenamefont {Fu}\ \emph {et~al.}(2024{\natexlab{b}})\citenamefont
  {Fu}, \citenamefont {Liu}, \citenamefont {Xu}, \citenamefont {Lee},\ and\
  \citenamefont {Sin~Ang}}]{Fu2024:arxiv}%
  \BibitemOpen
  \bibfield  {author} {\bibinfo {author} {\bibfnamefont {P.-H.}\ \bibnamefont
  {Fu}}, \bibinfo {author} {\bibfnamefont {J.-F.}\ \bibnamefont {Liu}},
  \bibinfo {author} {\bibfnamefont {Y.}~\bibnamefont {Xu}}, \bibinfo {author}
  {\bibfnamefont {C.~H.}\ \bibnamefont {Lee}},\ and\ \bibinfo {author}
  {\bibfnamefont {Y.}~\bibnamefont {Sin~Ang}},\ }\href@noop {} {\bibinfo
  {title} {Transverse cooper-pair rectifier}} (\bibinfo {year}
  {2024}{\natexlab{b}}),\ \Eprint {https://arxiv.org/abs/2405.04751}
  {arXiv:2405.04751} \BibitemShut {NoStop}%
\bibitem [{\citenamefont {Edelshtein}(1989)}]{Edelstein1989}%
  \BibitemOpen
  \bibfield  {author} {\bibinfo {author} {\bibfnamefont {V.~M.}\ \bibnamefont
  {Edelshtein}},\ }\bibfield  {title} {\bibinfo {title} {{Characteristics of
  the Cooper pairing in two-dimensional noncentrosymmetric electron systems}},\
  }\href {http://www.jetp.ras.ru/cgi-bin/dn/e_068_06_1244.pdf} {\bibfield
  {journal} {\bibinfo  {journal} {Sov. Phys. - JETP}\ }\textbf {\bibinfo
  {volume} {68}},\ \bibinfo {pages} {1244} (\bibinfo {year}
  {1989})}\BibitemShut {NoStop}%
\bibitem [{\citenamefont {Edelstein}(1996)}]{Edelstein1996}%
  \BibitemOpen
  \bibfield  {author} {\bibinfo {author} {\bibfnamefont {V.~M.}\ \bibnamefont
  {Edelstein}},\ }\bibfield  {title} {\bibinfo {title} {{The Ginzburg - Landau
  equation for superconductors of polar symmetry}},\ }\href
  {https://doi.org/10.1088/0953-8984/8/3/012} {\bibfield  {journal} {\bibinfo
  {journal} {Journal of Physics: Condensed Matter}\ }\textbf {\bibinfo {volume}
  {8}},\ \bibinfo {pages} {339} (\bibinfo {year} {1996})}\BibitemShut {NoStop}%
\bibitem [{\citenamefont {Daido}\ \emph {et~al.}(2022)\citenamefont {Daido},
  \citenamefont {Ikeda},\ and\ \citenamefont {Yanase}}]{Daido2021}%
  \BibitemOpen
  \bibfield  {author} {\bibinfo {author} {\bibfnamefont {A.}~\bibnamefont
  {Daido}}, \bibinfo {author} {\bibfnamefont {Y.}~\bibnamefont {Ikeda}},\ and\
  \bibinfo {author} {\bibfnamefont {Y.}~\bibnamefont {Yanase}},\ }\bibfield
  {title} {\bibinfo {title} {{Intrinsic Superconducting Diode Effect}},\ }\href
  {https://doi.org/10.1103/PhysRevLett.128.037001} {\bibfield  {journal}
  {\bibinfo  {journal} {Phys. Rev. Lett.}\ }\textbf {\bibinfo {volume} {128}},\
  \bibinfo {pages} {037001} (\bibinfo {year} {2022})}\BibitemShut {NoStop}%
\bibitem [{\citenamefont {Yuan}\ and\ \citenamefont {Fu}(2022)}]{Yuan2021}%
  \BibitemOpen
  \bibfield  {author} {\bibinfo {author} {\bibfnamefont {N.~F.~Q.}\
  \bibnamefont {Yuan}}\ and\ \bibinfo {author} {\bibfnamefont {L.}~\bibnamefont
  {Fu}},\ }\bibfield  {title} {\bibinfo {title} {{Supercurrent diode effect and
  finite-momentum superconductors}},\ }\href
  {https://doi.org/10.1073/pnas.2119548119} {\bibfield  {journal} {\bibinfo
  {journal} {Proc. Natl. Acad. Sci. USA}\ }\textbf {\bibinfo {volume} {119}},\
  \bibinfo {pages} {e2119548119} (\bibinfo {year} {2022})}\BibitemShut
  {NoStop}%
\bibitem [{\citenamefont {Smith}\ \emph {et~al.}(2021)\citenamefont {Smith},
  \citenamefont {Andreev},\ and\ \citenamefont {Spivak}}]{Smith2021}%
  \BibitemOpen
  \bibfield  {author} {\bibinfo {author} {\bibfnamefont {M.}~\bibnamefont
  {Smith}}, \bibinfo {author} {\bibfnamefont {A.~V.}\ \bibnamefont {Andreev}},\
  and\ \bibinfo {author} {\bibfnamefont {B.~Z.}\ \bibnamefont {Spivak}},\
  }\bibfield  {title} {\bibinfo {title} {{Giant magnetoconductivity in
  noncentrosymmetric superconductors}},\ }\href
  {https://doi.org/10.1103/PhysRevB.104.L220504} {\bibfield  {journal}
  {\bibinfo  {journal} {Phys. Rev. B}\ }\textbf {\bibinfo {volume} {104}},\
  \bibinfo {pages} {L220504} (\bibinfo {year} {2021})}\BibitemShut {NoStop}%
\bibitem [{\citenamefont {He}\ \emph {et~al.}(2022)\citenamefont {He},
  \citenamefont {Tanaka},\ and\ \citenamefont {Nagaosa}}]{He2022}%
  \BibitemOpen
  \bibfield  {author} {\bibinfo {author} {\bibfnamefont {J.~J.}\ \bibnamefont
  {He}}, \bibinfo {author} {\bibfnamefont {Y.}~\bibnamefont {Tanaka}},\ and\
  \bibinfo {author} {\bibfnamefont {N.}~\bibnamefont {Nagaosa}},\ }\bibfield
  {title} {\bibinfo {title} {{A phenomenological theory of superconductor
  diodes}},\ }\href {https://doi.org/10.1088/1367-2630/ac6766} {\bibfield
  {journal} {\bibinfo  {journal} {New~J.~Phys.}\ }\textbf {\bibinfo {volume}
  {24}},\ \bibinfo {pages} {053014} (\bibinfo {year} {2022})}\BibitemShut
  {NoStop}%
\bibitem [{\citenamefont {Scammell}\ \emph {et~al.}(2022)\citenamefont
  {Scammell}, \citenamefont {Li},\ and\ \citenamefont
  {Scheurer}}]{Scammell2022}%
  \BibitemOpen
  \bibfield  {author} {\bibinfo {author} {\bibfnamefont {H.~D.}\ \bibnamefont
  {Scammell}}, \bibinfo {author} {\bibfnamefont {J.~I.~A.}\ \bibnamefont
  {Li}},\ and\ \bibinfo {author} {\bibfnamefont {M.~S.}\ \bibnamefont
  {Scheurer}},\ }\bibfield  {title} {\bibinfo {title} {{Theory of zero-field
  superconducting diode effect in twisted trilayer graphene}},\ }\href
  {https://doi.org/10.1088/2053-1583/ac5b16} {\bibfield  {journal} {\bibinfo
  {journal} {2D~Mater.}\ }\textbf {\bibinfo {volume} {9}},\ \bibinfo {pages}
  {025027} (\bibinfo {year} {2022})}\BibitemShut {NoStop}%
\bibitem [{\citenamefont {Ili\ifmmode~\acute{c}\else \'{c}\fi{}}\ and\
  \citenamefont {Bergeret}(2022)}]{Ilic2022}%
  \BibitemOpen
  \bibfield  {author} {\bibinfo {author} {\bibfnamefont {S.}~\bibnamefont
  {Ili\ifmmode~\acute{c}\else \'{c}\fi{}}}\ and\ \bibinfo {author}
  {\bibfnamefont {F.~S.}\ \bibnamefont {Bergeret}},\ }\bibfield  {title}
  {\bibinfo {title} {{Theory of the Supercurrent Diode Effect in Rashba
  Superconductors with Arbitrary Disorder}},\ }\href
  {https://doi.org/10.1103/PhysRevLett.128.177001} {\bibfield  {journal}
  {\bibinfo  {journal} {Phys. Rev. Lett.}\ }\textbf {\bibinfo {volume} {128}},\
  \bibinfo {pages} {177001} (\bibinfo {year} {2022})}\BibitemShut {NoStop}%
\bibitem [{\citenamefont {de~Picoli}\ \emph {et~al.}(2023)\citenamefont
  {de~Picoli}, \citenamefont {Blood}, \citenamefont {Lyanda-Geller},\ and\
  \citenamefont {V\"ayrynen}}]{PicoliPRB:2023-Theory}%
  \BibitemOpen
  \bibfield  {author} {\bibinfo {author} {\bibfnamefont {T.}~\bibnamefont
  {de~Picoli}}, \bibinfo {author} {\bibfnamefont {Z.}~\bibnamefont {Blood}},
  \bibinfo {author} {\bibfnamefont {Y.}~\bibnamefont {Lyanda-Geller}},\ and\
  \bibinfo {author} {\bibfnamefont {J.~I.}\ \bibnamefont {V\"ayrynen}},\
  }\bibfield  {title} {\bibinfo {title} {{Superconducting diode effect in
  quasi-one-dimensional systems}},\ }\href
  {https://doi.org/10.1103/PhysRevB.107.224518} {\bibfield  {journal} {\bibinfo
   {journal} {Phys. Rev. B}\ }\textbf {\bibinfo {volume} {107}},\ \bibinfo
  {pages} {224518} (\bibinfo {year} {2023})}\BibitemShut {NoStop}%
\bibitem [{\citenamefont {Fuchs}\ \emph {et~al.}(2022)\citenamefont {Fuchs},
  \citenamefont {Kochan}, \citenamefont {Schmidt}, \citenamefont {H\"uttner},
  \citenamefont {Baumgartner}, \citenamefont {Reinhardt}, \citenamefont
  {Gronin}, \citenamefont {Gardner}, \citenamefont {Lindemann}, \citenamefont
  {Manfra}, \citenamefont {Strunk},\ and\ \citenamefont
  {Paradiso}}]{Fuchs2022}%
  \BibitemOpen
  \bibfield  {author} {\bibinfo {author} {\bibfnamefont {L.}~\bibnamefont
  {Fuchs}}, \bibinfo {author} {\bibfnamefont {D.}~\bibnamefont {Kochan}},
  \bibinfo {author} {\bibfnamefont {J.}~\bibnamefont {Schmidt}}, \bibinfo
  {author} {\bibfnamefont {N.}~\bibnamefont {H\"uttner}}, \bibinfo {author}
  {\bibfnamefont {C.}~\bibnamefont {Baumgartner}}, \bibinfo {author}
  {\bibfnamefont {S.}~\bibnamefont {Reinhardt}}, \bibinfo {author}
  {\bibfnamefont {S.}~\bibnamefont {Gronin}}, \bibinfo {author} {\bibfnamefont
  {G.~C.}\ \bibnamefont {Gardner}}, \bibinfo {author} {\bibfnamefont
  {T.}~\bibnamefont {Lindemann}}, \bibinfo {author} {\bibfnamefont {M.~J.}\
  \bibnamefont {Manfra}}, \bibinfo {author} {\bibfnamefont {C.}~\bibnamefont
  {Strunk}},\ and\ \bibinfo {author} {\bibfnamefont {N.}~\bibnamefont
  {Paradiso}},\ }\bibfield  {title} {\bibinfo {title} {{Anisotropic Vortex
  Squeezing in Synthetic Rashba Superconductors: A Manifestation of Lifshitz
  Invariants}},\ }\href {https://doi.org/10.1103/PhysRevX.12.041020} {\bibfield
   {journal} {\bibinfo  {journal} {Phys. Rev. X}\ }\textbf {\bibinfo {volume}
  {12}},\ \bibinfo {pages} {041020} (\bibinfo {year} {2022})}\BibitemShut
  {NoStop}%
\bibitem [{\citenamefont {Banerjee}\ \emph {et~al.}(2023)\citenamefont
  {Banerjee}, \citenamefont {Geier}, \citenamefont {Rahman}, \citenamefont
  {Thomas}, \citenamefont {Wang}, \citenamefont {Manfra}, \citenamefont
  {Flensberg},\ and\ \citenamefont {Marcus}}]{Banerjee2023phase}%
  \BibitemOpen
  \bibfield  {author} {\bibinfo {author} {\bibfnamefont {A.}~\bibnamefont
  {Banerjee}}, \bibinfo {author} {\bibfnamefont {M.}~\bibnamefont {Geier}},
  \bibinfo {author} {\bibfnamefont {M.~A.}\ \bibnamefont {Rahman}}, \bibinfo
  {author} {\bibfnamefont {C.}~\bibnamefont {Thomas}}, \bibinfo {author}
  {\bibfnamefont {T.}~\bibnamefont {Wang}}, \bibinfo {author} {\bibfnamefont
  {M.~J.}\ \bibnamefont {Manfra}}, \bibinfo {author} {\bibfnamefont
  {K.}~\bibnamefont {Flensberg}},\ and\ \bibinfo {author} {\bibfnamefont
  {C.~M.}\ \bibnamefont {Marcus}},\ }\bibfield  {title} {\bibinfo {title}
  {{Phase Asymmetry of Andreev Spectra from Cooper-Pair Momentum}},\ }\href
  {https://doi.org/10.1103/PhysRevLett.131.196301} {\bibfield  {journal}
  {\bibinfo  {journal} {Phys. Rev. Lett.}\ }\textbf {\bibinfo {volume} {131}},\
  \bibinfo {pages} {196301} (\bibinfo {year} {2023})}\BibitemShut {NoStop}%
\bibitem [{\citenamefont {Sundaresh}\ \emph {et~al.}(2023)\citenamefont
  {Sundaresh}, \citenamefont {V{\"a}yrynen}, \citenamefont {Lyanda-Geller},\
  and\ \citenamefont {Rokhinson}}]{Sundaresh2023:NatCom}%
  \BibitemOpen
  \bibfield  {author} {\bibinfo {author} {\bibfnamefont {A.}~\bibnamefont
  {Sundaresh}}, \bibinfo {author} {\bibfnamefont {J.~I.}\ \bibnamefont
  {V{\"a}yrynen}}, \bibinfo {author} {\bibfnamefont {Y.}~\bibnamefont
  {Lyanda-Geller}},\ and\ \bibinfo {author} {\bibfnamefont {L.~P.}\
  \bibnamefont {Rokhinson}},\ }\bibfield  {title} {\bibinfo {title}
  {{Diamagnetic mechanism of critical current non-reciprocity in multilayered
  superconductors}},\ }\href {https://doi.org/10.1038/s41467-023-36786-5}
  {\bibfield  {journal} {\bibinfo  {journal} {Nature Communications}\ }\textbf
  {\bibinfo {volume} {14}},\ \bibinfo {pages} {1628} (\bibinfo {year}
  {2023})}\BibitemShut {NoStop}%
\bibitem [{\citenamefont {Kochan}\ \emph {et~al.}(2023)\citenamefont {Kochan},
  \citenamefont {Costa}, \citenamefont {Zhumagulov},\ and\ \citenamefont
  {Žutić}}]{Kochan2023diode}%
  \BibitemOpen
  \bibfield  {author} {\bibinfo {author} {\bibfnamefont {D.}~\bibnamefont
  {Kochan}}, \bibinfo {author} {\bibfnamefont {A.}~\bibnamefont {Costa}},
  \bibinfo {author} {\bibfnamefont {I.}~\bibnamefont {Zhumagulov}},\ and\
  \bibinfo {author} {\bibfnamefont {I.}~\bibnamefont {Žutić}},\ }\href@noop
  {} {\bibinfo {title} {{Phenomenological Theory of the Supercurrent Diode
  Effect: The Lifshitz Invariant}}} (\bibinfo {year} {2023}),\ \Eprint
  {https://arxiv.org/abs/2303.11975} {arXiv:2303.11975 [cond-mat.supr-con]}
  \BibitemShut {NoStop}%
\bibitem [{\citenamefont {Costa}\ \emph
  {et~al.}(2023{\natexlab{b}})\citenamefont {Costa}, \citenamefont {Fabian},\
  and\ \citenamefont {Kochan}}]{Costa2023:PRB-Theory}%
  \BibitemOpen
  \bibfield  {author} {\bibinfo {author} {\bibfnamefont {A.}~\bibnamefont
  {Costa}}, \bibinfo {author} {\bibfnamefont {J.}~\bibnamefont {Fabian}},\ and\
  \bibinfo {author} {\bibfnamefont {D.}~\bibnamefont {Kochan}},\ }\bibfield
  {title} {\bibinfo {title} {{Microscopic study of the Josephson supercurrent
  diode effect in Josephson junctions based on two-dimensional electron gas}},\
  }\href {https://doi.org/10.1103/PhysRevB.108.054522} {\bibfield  {journal}
  {\bibinfo  {journal} {Phys. Rev. B}\ }\textbf {\bibinfo {volume} {108}},\
  \bibinfo {pages} {054522} (\bibinfo {year} {2023}{\natexlab{b}})}\BibitemShut
  {NoStop}%
\bibitem [{\citenamefont {Mineev}\ and\ \citenamefont
  {Samokhin}(1994)}]{Mineev1994}%
  \BibitemOpen
  \bibfield  {author} {\bibinfo {author} {\bibfnamefont {V.}~\bibnamefont
  {Mineev}}\ and\ \bibinfo {author} {\bibfnamefont {K.}~\bibnamefont
  {Samokhin}},\ }\bibfield  {title} {\bibinfo {title} {{Helical phases in
  superconductors}},\ }\href
  {http://www.jetp.ras.ru/cgi-bin/dn/e_078_03_0401.pdf} {\bibfield  {journal}
  {\bibinfo  {journal} {JETP}\ }\textbf {\bibinfo {volume} {78}},\ \bibinfo
  {pages} {401} (\bibinfo {year} {1994})}\BibitemShut {NoStop}%
\bibitem [{\citenamefont {Dimitrova}\ and\ \citenamefont
  {Feigel'man}(2007)}]{Dimitrova2007}%
  \BibitemOpen
  \bibfield  {author} {\bibinfo {author} {\bibfnamefont {O.}~\bibnamefont
  {Dimitrova}}\ and\ \bibinfo {author} {\bibfnamefont {M.~V.}\ \bibnamefont
  {Feigel'man}},\ }\bibfield  {title} {\bibinfo {title} {{Theory of a
  two-dimensional superconductor with broken inversion symmetry}},\ }\href
  {https://doi.org/10.1103/PhysRevB.76.014522} {\bibfield  {journal} {\bibinfo
  {journal} {Phys. Rev. B}\ }\textbf {\bibinfo {volume} {76}},\ \bibinfo
  {pages} {014522} (\bibinfo {year} {2007})}\BibitemShut {NoStop}%
\bibitem [{\citenamefont {Buzdin}(2008)}]{Buzdin2008}%
  \BibitemOpen
  \bibfield  {author} {\bibinfo {author} {\bibfnamefont {A.}~\bibnamefont
  {Buzdin}},\ }\bibfield  {title} {\bibinfo {title} {{Direct Coupling Between
  Magnetism and Superconducting Current in the Josephson
  ${\ensuremath{\varphi}}_{0}$ Junction}},\ }\href
  {https://doi.org/10.1103/PhysRevLett.101.107005} {\bibfield  {journal}
  {\bibinfo  {journal} {Phys. Rev. Lett.}\ }\textbf {\bibinfo {volume} {101}},\
  \bibinfo {pages} {107005} (\bibinfo {year} {2008})}\BibitemShut {NoStop}%
\bibitem [{\citenamefont {Mineev}\ and\ \citenamefont
  {Sigrist}(2012)}]{Mineev2012}%
  \BibitemOpen
  \bibfield  {author} {\bibinfo {author} {\bibfnamefont {V.~P.}\ \bibnamefont
  {Mineev}}\ and\ \bibinfo {author} {\bibfnamefont {M.}~\bibnamefont
  {Sigrist}},\ }\bibfield  {title} {\bibinfo {title} {{Basic Theory of
  Superconductivity in Metals Without Inversion Center}},\ }in\ \href
  {https://doi.org/10.1007/978-3-642-24624-1_4} {\emph {\bibinfo {booktitle}
  {{Non-Centrosymmetric Superconductors}}}},\ \bibinfo {editor} {edited by\
  \bibinfo {editor} {\bibfnamefont {E.}~\bibnamefont {Bauer}}\ and\ \bibinfo
  {editor} {\bibfnamefont {M.}~\bibnamefont {Sigrist}}}\ (\bibinfo  {publisher}
  {Springer Berlin Heidelberg},\ \bibinfo {year} {2012})\ pp.\ \bibinfo {pages}
  {129--154}\BibitemShut {NoStop}%
\bibitem [{\citenamefont {Costa}\ \emph {et~al.}(2018)\citenamefont {Costa},
  \citenamefont {Fabian},\ and\ \citenamefont {Kochan}}]{Costa2018}%
  \BibitemOpen
  \bibfield  {author} {\bibinfo {author} {\bibfnamefont {A.}~\bibnamefont
  {Costa}}, \bibinfo {author} {\bibfnamefont {J.}~\bibnamefont {Fabian}},\ and\
  \bibinfo {author} {\bibfnamefont {D.}~\bibnamefont {Kochan}},\ }\bibfield
  {title} {\bibinfo {title} {Connection between zero-energy yu-shiba-rusinov
  states and $0\text{\ensuremath{-}}\ensuremath{\pi}$ transitions in magnetic
  josephson junctions},\ }\href {https://doi.org/10.1103/PhysRevB.98.134511}
  {\bibfield  {journal} {\bibinfo  {journal} {Phys. Rev. B}\ }\textbf {\bibinfo
  {volume} {98}},\ \bibinfo {pages} {134511} (\bibinfo {year}
  {2018})}\BibitemShut {NoStop}%
\bibitem [{\citenamefont {Dolcini}\ \emph {et~al.}(2015)\citenamefont
  {Dolcini}, \citenamefont {Houzet},\ and\ \citenamefont
  {Meyer}}]{Dolcini2015:PRB}%
  \BibitemOpen
  \bibfield  {author} {\bibinfo {author} {\bibfnamefont {F.}~\bibnamefont
  {Dolcini}}, \bibinfo {author} {\bibfnamefont {M.}~\bibnamefont {Houzet}},\
  and\ \bibinfo {author} {\bibfnamefont {J.~S.}\ \bibnamefont {Meyer}},\
  }\bibfield  {title} {\bibinfo {title} {Topological josephson
  ${\ensuremath{\phi}}_{0}$ junctions},\ }\href
  {https://doi.org/10.1103/PhysRevB.92.035428} {\bibfield  {journal} {\bibinfo
  {journal} {Phys. Rev. B}\ }\textbf {\bibinfo {volume} {92}},\ \bibinfo
  {pages} {035428} (\bibinfo {year} {2015})}\BibitemShut {NoStop}%
\bibitem [{\citenamefont {Scharf}\ \emph {et~al.}(2021)\citenamefont {Scharf},
  \citenamefont {Braggio}, \citenamefont {Strambini}, \citenamefont
  {Giazotto},\ and\ \citenamefont {Hankiewicz}}]{Scharf2021:PRR}%
  \BibitemOpen
  \bibfield  {author} {\bibinfo {author} {\bibfnamefont {B.}~\bibnamefont
  {Scharf}}, \bibinfo {author} {\bibfnamefont {A.}~\bibnamefont {Braggio}},
  \bibinfo {author} {\bibfnamefont {E.}~\bibnamefont {Strambini}}, \bibinfo
  {author} {\bibfnamefont {F.}~\bibnamefont {Giazotto}},\ and\ \bibinfo
  {author} {\bibfnamefont {E.~M.}\ \bibnamefont {Hankiewicz}},\ }\bibfield
  {title} {\bibinfo {title} {Thermodynamics in topological josephson
  junctions},\ }\href {https://doi.org/10.1103/PhysRevResearch.3.033062}
  {\bibfield  {journal} {\bibinfo  {journal} {Phys. Rev. Res.}\ }\textbf
  {\bibinfo {volume} {3}},\ \bibinfo {pages} {033062} (\bibinfo {year}
  {2021})}\BibitemShut {NoStop}%
\bibitem [{\citenamefont {Tkachov}\ \emph {et~al.}(2015)\citenamefont
  {Tkachov}, \citenamefont {Burset}, \citenamefont {Trauzettel},\ and\
  \citenamefont {Hankiewicz}}]{Tkachov2015:PRB}%
  \BibitemOpen
  \bibfield  {author} {\bibinfo {author} {\bibfnamefont {G.}~\bibnamefont
  {Tkachov}}, \bibinfo {author} {\bibfnamefont {P.}~\bibnamefont {Burset}},
  \bibinfo {author} {\bibfnamefont {B.}~\bibnamefont {Trauzettel}},\ and\
  \bibinfo {author} {\bibfnamefont {E.~M.}\ \bibnamefont {Hankiewicz}},\
  }\bibfield  {title} {\bibinfo {title} {Quantum interference of edge
  supercurrents in a two-dimensional topological insulator},\ }\href
  {https://doi.org/10.1103/PhysRevB.92.045408} {\bibfield  {journal} {\bibinfo
  {journal} {Phys. Rev. B}\ }\textbf {\bibinfo {volume} {92}},\ \bibinfo
  {pages} {045408} (\bibinfo {year} {2015})}\BibitemShut {NoStop}%
\bibitem [{\citenamefont {Tkachov}(2019)}]{Tkachov2019:PRB}%
  \BibitemOpen
  \bibfield  {author} {\bibinfo {author} {\bibfnamefont {G.}~\bibnamefont
  {Tkachov}},\ }\bibfield  {title} {\bibinfo {title} {Chiral current-phase
  relation of topological josephson junctions: A signature of the
  $4\ensuremath{\pi}$-periodic josephson effect},\ }\href
  {https://doi.org/10.1103/PhysRevB.100.035403} {\bibfield  {journal} {\bibinfo
   {journal} {Phys. Rev. B}\ }\textbf {\bibinfo {volume} {100}},\ \bibinfo
  {pages} {035403} (\bibinfo {year} {2019})}\BibitemShut {NoStop}%
\bibitem [{Note1()}]{Note1}%
  \BibitemOpen
  \bibinfo {note} {For the bottom edge, there is a minimum at $\gamma
  =1/\protect \sqrt {1+\pi ^2/16}$ with a value of $Q^b=-\protect \sqrt {16/\pi
  ^2+1}+1$.}\BibitemShut {Stop}%
\bibitem [{\citenamefont {Ioselevich}\ and\ \citenamefont
  {Feigel'man}(2011)}]{Ioselevich2011:PRL}%
  \BibitemOpen
  \bibfield  {author} {\bibinfo {author} {\bibfnamefont {P.~A.}\ \bibnamefont
  {Ioselevich}}\ and\ \bibinfo {author} {\bibfnamefont {M.~V.}\ \bibnamefont
  {Feigel'man}},\ }\bibfield  {title} {\bibinfo {title} {Anomalous josephson
  current via majorana bound states in topological insulators},\ }\href
  {https://doi.org/10.1103/PhysRevLett.106.077003} {\bibfield  {journal}
  {\bibinfo  {journal} {Phys. Rev. Lett.}\ }\textbf {\bibinfo {volume} {106}},\
  \bibinfo {pages} {077003} (\bibinfo {year} {2011})}\BibitemShut {NoStop}%
\bibitem [{\citenamefont {Beenakker}\ \emph {et~al.}(2013)\citenamefont
  {Beenakker}, \citenamefont {Pikulin}, \citenamefont {Hyart}, \citenamefont
  {Schomerus},\ and\ \citenamefont {Dahlhaus}}]{Beenakker2013:PRL}%
  \BibitemOpen
  \bibfield  {author} {\bibinfo {author} {\bibfnamefont {C.~W.~J.}\
  \bibnamefont {Beenakker}}, \bibinfo {author} {\bibfnamefont {D.~I.}\
  \bibnamefont {Pikulin}}, \bibinfo {author} {\bibfnamefont {T.}~\bibnamefont
  {Hyart}}, \bibinfo {author} {\bibfnamefont {H.}~\bibnamefont {Schomerus}},\
  and\ \bibinfo {author} {\bibfnamefont {J.~P.}\ \bibnamefont {Dahlhaus}},\
  }\bibfield  {title} {\bibinfo {title} {Fermion-parity anomaly of the critical
  supercurrent in the quantum spin-hall effect},\ }\href
  {https://doi.org/10.1103/PhysRevLett.110.017003} {\bibfield  {journal}
  {\bibinfo  {journal} {Phys. Rev. Lett.}\ }\textbf {\bibinfo {volume} {110}},\
  \bibinfo {pages} {017003} (\bibinfo {year} {2013})}\BibitemShut {NoStop}%
\bibitem [{Note2()}]{Note2}%
  \BibitemOpen
  \bibinfo {note} {Assuming a quasi-particle density of states $\rho
  _S(\epsilon )=2/(\pi E_S)|\epsilon |\Theta \left (\epsilon ^2-\Delta ^2\right
  )/\protect \sqrt {\epsilon ^2-\Delta ^2}$ from the superconducting
  electrodes, one can compute its contribution to Eq.~(\ref
  {eq:FreeEnergyParity}) via $J_S(T)=\DOTSI \intop \ilimits@ \limits _0^\infty
  \protect \ensuremath {\protect \mathrm {d}}\epsilon \rho _S(\epsilon )\ln
  \protect \Bigl [\tanh \protect \Bigl (\protect \frac {\epsilon }{2k_{\protect
  \text {B}}T}\protect \Bigr )\protect \Bigr ]$, which yields Eq.~(\ref
  {eq:Js}), as detailed in Refs.~\protect \rev@citealp
  {Beenakker2013:PRL,Scharf2021:PRR}.}\BibitemShut {Stop}%
\bibitem [{\citenamefont {Bernevig}\ \emph {et~al.}(2006)\citenamefont
  {Bernevig}, \citenamefont {Hughes},\ and\ \citenamefont {Zhang}}]{BHZ2006:S}%
  \BibitemOpen
  \bibfield  {author} {\bibinfo {author} {\bibfnamefont {B.~A.}\ \bibnamefont
  {Bernevig}}, \bibinfo {author} {\bibfnamefont {T.~L.}\ \bibnamefont
  {Hughes}},\ and\ \bibinfo {author} {\bibfnamefont {S.-C.}\ \bibnamefont
  {Zhang}},\ }\bibfield  {title} {\bibinfo {title} {Quantum spin hall effect
  and topological phase transition in hgte quantum wells},\ }\href
  {https://doi.org/10.1126/science.1133734} {\bibfield  {journal} {\bibinfo
  {journal} {Science}\ }\textbf {\bibinfo {volume} {314}},\ \bibinfo {pages}
  {1757} (\bibinfo {year} {2006})}\BibitemShut {NoStop}%
\bibitem [{\citenamefont {K\"{o}nig}\ \emph {et~al.}(2007)\citenamefont
  {K\"{o}nig}, \citenamefont {Wiedmann}, \citenamefont {Br\"{u}ne},
  \citenamefont {Roth}, \citenamefont {Buhmann}, \citenamefont {Molenkamp},
  \citenamefont {Qi},\ and\ \citenamefont {Zhang}}]{Koenig2007:S}%
  \BibitemOpen
  \bibfield  {author} {\bibinfo {author} {\bibfnamefont {M.}~\bibnamefont
  {K\"{o}nig}}, \bibinfo {author} {\bibfnamefont {S.}~\bibnamefont {Wiedmann}},
  \bibinfo {author} {\bibfnamefont {C.}~\bibnamefont {Br\"{u}ne}}, \bibinfo
  {author} {\bibfnamefont {A.}~\bibnamefont {Roth}}, \bibinfo {author}
  {\bibfnamefont {H.}~\bibnamefont {Buhmann}}, \bibinfo {author} {\bibfnamefont
  {L.~W.}\ \bibnamefont {Molenkamp}}, \bibinfo {author} {\bibfnamefont {X.-L.}\
  \bibnamefont {Qi}},\ and\ \bibinfo {author} {\bibfnamefont {S.-C.}\
  \bibnamefont {Zhang}},\ }\bibfield  {title} {\bibinfo {title} {Quantum spin
  hall insulator state in hgte quantum wells},\ }\href
  {https://doi.org/10.1126/science.1148047} {\bibfield  {journal} {\bibinfo
  {journal} {Science}\ }\textbf {\bibinfo {volume} {318}},\ \bibinfo {pages}
  {766} (\bibinfo {year} {2007})}\BibitemShut {NoStop}%
\bibitem [{\citenamefont {Sengupta}\ \emph {et~al.}(2013)\citenamefont
  {Sengupta}, \citenamefont {Kubis}, \citenamefont {Tan}, \citenamefont
  {Povolotskyi},\ and\ \citenamefont {Klimeck}}]{Sengupta2013:JAP}%
  \BibitemOpen
  \bibfield  {author} {\bibinfo {author} {\bibfnamefont {P.}~\bibnamefont
  {Sengupta}}, \bibinfo {author} {\bibfnamefont {T.}~\bibnamefont {Kubis}},
  \bibinfo {author} {\bibfnamefont {Y.}~\bibnamefont {Tan}}, \bibinfo {author}
  {\bibfnamefont {M.}~\bibnamefont {Povolotskyi}},\ and\ \bibinfo {author}
  {\bibfnamefont {G.}~\bibnamefont {Klimeck}},\ }\bibfield  {title} {\bibinfo
  {title} {{Design principles for HgTe based topological insulator devices}},\
  }\href {https://doi.org/10.1063/1.4813877} {\bibfield  {journal} {\bibinfo
  {journal} {Journal of Applied Physics}\ }\textbf {\bibinfo {volume} {114}},\
  \bibinfo {pages} {043702} (\bibinfo {year} {2013})}\BibitemShut {NoStop}%
\bibitem [{\citenamefont {Topalovic}\ \emph {et~al.}(2020)\citenamefont
  {Topalovic}, \citenamefont {Arsoski}, \citenamefont {Tadic},\ and\
  \citenamefont {Peeters}}]{Topalovic2020:JAP}%
  \BibitemOpen
  \bibfield  {author} {\bibinfo {author} {\bibfnamefont {D.~B.}\ \bibnamefont
  {Topalovic}}, \bibinfo {author} {\bibfnamefont {V.~V.}\ \bibnamefont
  {Arsoski}}, \bibinfo {author} {\bibfnamefont {M.~Z.}\ \bibnamefont {Tadic}},\
  and\ \bibinfo {author} {\bibfnamefont {F.~M.}\ \bibnamefont {Peeters}},\
  }\bibfield  {title} {\bibinfo {title} {{Asymmetric versus symmetric
  HgTe/Cd$_x$Hg$_{1-x}$Te double quantum wells: Bandgap tuning without electric
  field}},\ }\href {https://doi.org/10.1063/5.0016069} {\bibfield  {journal}
  {\bibinfo  {journal} {Journal of Applied Physics}\ }\textbf {\bibinfo
  {volume} {128}},\ \bibinfo {pages} {064301} (\bibinfo {year}
  {2020})}\BibitemShut {NoStop}%
\bibitem [{\citenamefont {Hart}\ \emph {et~al.}(2017)\citenamefont {Hart},
  \citenamefont {Ren}, \citenamefont {Kosowsky}, \citenamefont {Ben-Shach},
  \citenamefont {Leubner}, \citenamefont {Br{\"u}ne}, \citenamefont {Buhmann},
  \citenamefont {Molenkamp}, \citenamefont {Halperin},\ and\ \citenamefont
  {Yacoby}}]{Hart2017:NP}%
  \BibitemOpen
  \bibfield  {author} {\bibinfo {author} {\bibfnamefont {S.}~\bibnamefont
  {Hart}}, \bibinfo {author} {\bibfnamefont {H.}~\bibnamefont {Ren}}, \bibinfo
  {author} {\bibfnamefont {M.}~\bibnamefont {Kosowsky}}, \bibinfo {author}
  {\bibfnamefont {G.}~\bibnamefont {Ben-Shach}}, \bibinfo {author}
  {\bibfnamefont {P.}~\bibnamefont {Leubner}}, \bibinfo {author} {\bibfnamefont
  {C.}~\bibnamefont {Br{\"u}ne}}, \bibinfo {author} {\bibfnamefont
  {H.}~\bibnamefont {Buhmann}}, \bibinfo {author} {\bibfnamefont {L.~W.}\
  \bibnamefont {Molenkamp}}, \bibinfo {author} {\bibfnamefont {B.~I.}\
  \bibnamefont {Halperin}},\ and\ \bibinfo {author} {\bibfnamefont
  {A.}~\bibnamefont {Yacoby}},\ }\bibfield  {title} {\bibinfo {title}
  {Controlled finite momentum pairing and spatially varying order parameter in
  proximitized hgte quantum wells},\ }\href {https://doi.org/10.1038/nphys3877}
  {\bibfield  {journal} {\bibinfo  {journal} {Nat. Phys.}\ }\textbf {\bibinfo
  {volume} {13}},\ \bibinfo {pages} {87} (\bibinfo {year} {2017})}\BibitemShut
  {NoStop}%
\bibitem [{\citenamefont {Ren}\ \emph {et~al.}(2019)\citenamefont {Ren},
  \citenamefont {Pientka}, \citenamefont {Hart}, \citenamefont {Pierce},
  \citenamefont {Kosowsky}, \citenamefont {Lunczer}, \citenamefont {Schlereth},
  \citenamefont {Scharf}, \citenamefont {Hankiewicz}, \citenamefont
  {Molenkamp}, \citenamefont {Halperin},\ and\ \citenamefont
  {Yacoby}}]{Ren2019:N}%
  \BibitemOpen
  \bibfield  {author} {\bibinfo {author} {\bibfnamefont {H.}~\bibnamefont
  {Ren}}, \bibinfo {author} {\bibfnamefont {F.}~\bibnamefont {Pientka}},
  \bibinfo {author} {\bibfnamefont {S.}~\bibnamefont {Hart}}, \bibinfo {author}
  {\bibfnamefont {A.~T.}\ \bibnamefont {Pierce}}, \bibinfo {author}
  {\bibfnamefont {M.}~\bibnamefont {Kosowsky}}, \bibinfo {author}
  {\bibfnamefont {L.}~\bibnamefont {Lunczer}}, \bibinfo {author} {\bibfnamefont
  {R.}~\bibnamefont {Schlereth}}, \bibinfo {author} {\bibfnamefont
  {B.}~\bibnamefont {Scharf}}, \bibinfo {author} {\bibfnamefont {E.~M.}\
  \bibnamefont {Hankiewicz}}, \bibinfo {author} {\bibfnamefont {L.~W.}\
  \bibnamefont {Molenkamp}}, \bibinfo {author} {\bibfnamefont {B.~I.}\
  \bibnamefont {Halperin}},\ and\ \bibinfo {author} {\bibfnamefont
  {A.}~\bibnamefont {Yacoby}},\ }\bibfield  {title} {\bibinfo {title}
  {Topological superconductivity in a phase-controlled josephson junction},\
  }\href {https://doi.org/10.1038/s41586-019-1148-9} {\bibfield  {journal}
  {\bibinfo  {journal} {Nature}\ }\textbf {\bibinfo {volume} {569}},\ \bibinfo
  {pages} {93} (\bibinfo {year} {2019})}\BibitemShut {NoStop}%
\bibitem [{Note3()}]{Note3}%
  \BibitemOpen
  \bibinfo {note} {Assuming, for example, the system from Ref.~\protect
  \rev@citealp {Ren2019:N} with an induced superconducting gap of $\Delta
  \approx 64$ $\mu $eV, a Fermi velocity of $v_F=5\times 10^5$ m/s and a width
  of the junction of $\protect \mathcal {W}_S=1$ $\mu $m, we obtain $\gamma
  \approx 3.9$ for $B=1$ mT, showing that no excessively large magnetic fields
  are necessary to observe a sizeable SDE.}\BibitemShut {Stop}%
\bibitem [{\citenamefont {Lutchyn}\ \emph {et~al.}(2010)\citenamefont
  {Lutchyn}, \citenamefont {Sau},\ and\ \citenamefont
  {Das~Sarma}}]{Lutchyn2010:PRL}%
  \BibitemOpen
  \bibfield  {author} {\bibinfo {author} {\bibfnamefont {R.~M.}\ \bibnamefont
  {Lutchyn}}, \bibinfo {author} {\bibfnamefont {J.~D.}\ \bibnamefont {Sau}},\
  and\ \bibinfo {author} {\bibfnamefont {S.}~\bibnamefont {Das~Sarma}},\
  }\bibfield  {title} {\bibinfo {title} {Majorana fermions and a topological
  phase transition in semiconductor-superconductor heterostructures},\ }\href
  {https://doi.org/10.1103/PhysRevLett.105.077001} {\bibfield  {journal}
  {\bibinfo  {journal} {Phys. Rev. Lett.}\ }\textbf {\bibinfo {volume} {105}},\
  \bibinfo {pages} {077001} (\bibinfo {year} {2010})}\BibitemShut {NoStop}%
\bibitem [{\citenamefont {Chiu}\ and\ \citenamefont
  {Das~Sarma}(2019)}]{Chiu2019:PRB}%
  \BibitemOpen
  \bibfield  {author} {\bibinfo {author} {\bibfnamefont {C.-K.}\ \bibnamefont
  {Chiu}}\ and\ \bibinfo {author} {\bibfnamefont {S.}~\bibnamefont
  {Das~Sarma}},\ }\bibfield  {title} {\bibinfo {title} {Fractional josephson
  effect with and without majorana zero modes},\ }\href
  {https://doi.org/10.1103/PhysRevB.99.035312} {\bibfield  {journal} {\bibinfo
  {journal} {Phys. Rev. B}\ }\textbf {\bibinfo {volume} {99}},\ \bibinfo
  {pages} {035312} (\bibinfo {year} {2019})}\BibitemShut {NoStop}%
\bibitem [{\citenamefont {Rainis}\ and\ \citenamefont
  {Loss}(2012)}]{Rainis2012:PRB}%
  \BibitemOpen
  \bibfield  {author} {\bibinfo {author} {\bibfnamefont {D.}~\bibnamefont
  {Rainis}}\ and\ \bibinfo {author} {\bibfnamefont {D.}~\bibnamefont {Loss}},\
  }\bibfield  {title} {\bibinfo {title} {Majorana qubit decoherence by
  quasiparticle poisoning},\ }\href
  {https://doi.org/10.1103/PhysRevB.85.174533} {\bibfield  {journal} {\bibinfo
  {journal} {Phys. Rev. B}\ }\textbf {\bibinfo {volume} {85}},\ \bibinfo
  {pages} {174533} (\bibinfo {year} {2012})}\BibitemShut {NoStop}%
\bibitem [{\citenamefont {Virtanen}\ and\ \citenamefont
  {Recher}(2013)}]{Virtanen2013:PRB}%
  \BibitemOpen
  \bibfield  {author} {\bibinfo {author} {\bibfnamefont {P.}~\bibnamefont
  {Virtanen}}\ and\ \bibinfo {author} {\bibfnamefont {P.}~\bibnamefont
  {Recher}},\ }\bibfield  {title} {\bibinfo {title} {Microwave spectroscopy of
  josephson junctions in topological superconductors},\ }\href
  {https://doi.org/10.1103/PhysRevB.88.144507} {\bibfield  {journal} {\bibinfo
  {journal} {Phys. Rev. B}\ }\textbf {\bibinfo {volume} {88}},\ \bibinfo
  {pages} {144507} (\bibinfo {year} {2013})}\BibitemShut {NoStop}%
\bibitem [{\citenamefont {Frombach}\ and\ \citenamefont
  {Recher}(2020)}]{Frombach2020:PRB}%
  \BibitemOpen
  \bibfield  {author} {\bibinfo {author} {\bibfnamefont {D.}~\bibnamefont
  {Frombach}}\ and\ \bibinfo {author} {\bibfnamefont {P.}~\bibnamefont
  {Recher}},\ }\bibfield  {title} {\bibinfo {title} {Quasiparticle poisoning
  effects on the dynamics of topological josephson junctions},\ }\href
  {https://doi.org/10.1103/PhysRevB.101.115304} {\bibfield  {journal} {\bibinfo
   {journal} {Phys. Rev. B}\ }\textbf {\bibinfo {volume} {101}},\ \bibinfo
  {pages} {115304} (\bibinfo {year} {2020})}\BibitemShut {NoStop}%
\bibitem [{\citenamefont {M\"{u}ck}\ and\ \citenamefont
  {McDermott}(2010)}]{Mueck2010:SST}%
  \BibitemOpen
  \bibfield  {author} {\bibinfo {author} {\bibfnamefont {M.}~\bibnamefont
  {M\"{u}ck}}\ and\ \bibinfo {author} {\bibfnamefont {R.}~\bibnamefont
  {McDermott}},\ }\bibfield  {title} {\bibinfo {title} {Radio-frequency
  amplifiers based on dc {SQUIDs}},\ }\href
  {https://doi.org/10.1088/0953-2048/23/9/093001} {\bibfield  {journal}
  {\bibinfo  {journal} {Supercond. Sci. Technol.}\ }\textbf {\bibinfo {volume}
  {23}},\ \bibinfo {pages} {093001} (\bibinfo {year} {2010})}\BibitemShut
  {NoStop}%
\end{thebibliography}%

\end{document}